\documentclass[cp1251
               ]{jetp} 
\twocolumn 

\begin{document}
{\English

\title{Topological Superfluids}

\setaffiliation1{Low Temperature Laboratory, Aalto University,  \\ P.O. Box 15100, FI-00076 Aalto, Finland}

\setaffiliation2{Landau Institute for Theoretical Physics, \\  acad. Semyonov av., 1a, 142432,
Chernogolovka, Russia}

\setauthor{G.~E.}{Volovik}{12}
\email{volovik@boojum.hut.fi}
 
\rtitle{Landau Institute}
\rauthor{G.~E. Volovik}

\maketitle




There are many topological faces of the superfluid phases of $^3$He. These superfluids contain various topological defects and textures. The momentum space topology of these superfluids is also nontrivial,  as well as the topology in the combined (${\bf p},{\bf r})$ phase space, giving rise to topologically protected Dirac, Weyl and Majorana fermions living in bulk, on the surface and within the topological objects.
The nontrivial topology lead to different types of anomalies, which extended in many different directions 
the Landau-Khalatninkov theory of superfluidity. 

\section{Introduction}

Superfluid phases of $^3$He discovered in 1972\cite{OsheroffRichardsonLee1972} opened the new area of the application of topological methods to condensed matter systems. Due to the multi-component order parameter which characterizes the broken symmetries in these phases, there are many inhomogeneous objects -- textures and defects in the order parameter field -- which are protected by topology and are characterized by topological charges. Among them there are quantized vortices, skyrmions and merons, solitons and vortex sheets, monopoles and boojums, Alice strings, Kibble walls terminated by Alice strings, spin vortices with soliton tails, etc.  Some of them have been experimentally identified and investigated\cite{SalomaaVolovik1987,Finne2006,Autti2016,Makinen2018},  the others are still waiting for their creation and detection.

The real-space topology, which is responsible for the topological stability of textures and defects, has been later extended to the  topology in momentum space, which governs the topologically protected properties of the ground state of these systems. This includes in particular the existence of the topologically stable nodes in the fermionic spectrum in bulk and/or on the surface of superfluids\cite{Volovik1987,SalomaaVolovik1988}. It appeared that the superfluid phases of liquid $^3$He serve as the clean examples of the topological matter, where 
 the momentum-space topology plays an important role in the properties of these phases\cite{HasanKane2010,Xiao-LiangQi2011,Schnyder2008,Kitaev2009}. The further natural extension
was  to the combined phase-space topology \cite{Grinevich1988},  which in particular describes the robust properties of the spectrum of fermionic states  
localized on topological defects.

In bulk liquid $^3$He there are two topologically different superfluid phases, $^3$He-A and $^3$He-B \cite{VollhardtWoelfle2013}.   One is  the chiral superfluid $^3$He-A with topologically protected Weyl points in the quasiparticle spectrum. In the vicinity  of the Weyl points, quasiparticles obey the Weyl equation and behave as Weyl fermions, with all the accompanying effects such as chiral 
anomaly\cite{Volovik1986a}, chiral magnetic effect (CME), chiral vortical effect (CVE)\cite{Volovik2003},  etc. The Adler-Bell-Jackiw equation, which describes the anomalous production of fermions from the vacuum \cite{Adler1969,BellJackiw1969,Adler2005}, has been  verified in experiments with skyrmions in $^3$He-A \cite{BevanNature1997}.
Weyl fermions have been  reported to exist in the topological semiconductors, which got the name Weyl semimetals
\cite{Herring1937,Abrikosov1971,Abrikosov1972,Weng2015,Huang2015,Lv2015,Xu2015,Lu2015},
see reviews \cite{Burkov2018,Bernevig2018,Armitage2018}. The possible manifestation of the chiral anomaly in these materials is under discussion \cite{Spivak2016}. 

Another phase is the fully gapped  time reversal invariant superfluid $^3$He-B. It has topologically protected gapless Majorana fermions living on the surface (see reviews \cite{Mizushima2015,Mizushima2016} on the momentum space 
topology in superfluid $^3$He).

The polar phase of $^3$He has been stabilized in $^3$He confined in the nematically ordered aerogel \cite{Dmitriev2015,HalperinParpiaSauls2018,Halperin2019,Fomin2018}. 
It is the time reversal invariant superfluid, which contains Dirac nodal ring in the fermionic spectrum \cite{Volovik2018a}.

\section{Topological defects in real space}

\begin{figure}
 \includegraphics[width=0.5\textwidth]{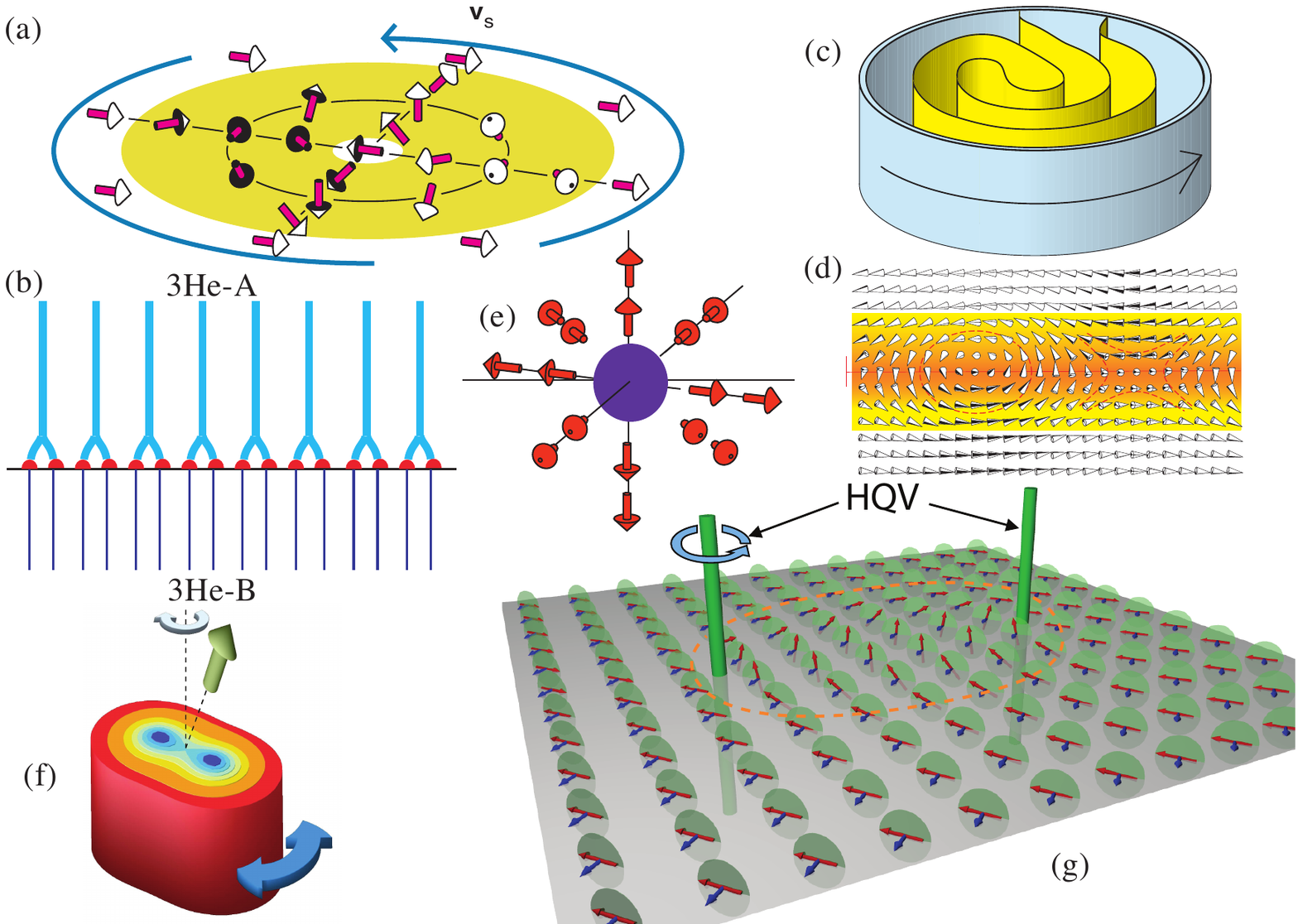}
 \caption{Some topological oblects in topological superfluids. (a) Vortex-skyrmions with two quanta of circulation 
(${\cal N}=2$) in the chiral superfluid $^3$He-A (see also Fig. \ref{Skyrmion2merons}b).
(b) Interface between $^3$He-A and $^3$He-B in rotation:
lattice of skyrmions with ${\cal N}=2$ in $^3$He-A transforms to the lattice of singly quantized vortices (${\cal N}=1$) in $^3$He-B. Each skyrmion with ${\cal N}=2$ splits into two merons with ${\cal N}=1$ (Mermin-Ho vortices). The Mermin-Ho vortex terminates on the boojum -- the surface singularity which is the analog of the
Dirac monopole  terminating the
$^3$He-B string (the ${\cal N}=1$ vortex).
(c) Vortex sheet in $^3$He-A.
(d) Elements of the vortex sheet: merons  with ${\cal N}=1$ within topological soliton in $^3$He-A.
(e) Hedgehog-monopole in the $\hat{\bf d}$-vector field in $^3$He-A. (f) Singly quantized vortex (${\cal N}=1$) with spontaneously broken axial symmetry in $^3$He-B as a pair of half-quantum vortices (${\cal N}=1/2$) connected by non-topological (dark) soliton.\cite{Volovik1990,SilaevThuneberg2015} Such vortex has been identified due to Goldstone mode associated with 
the spontaneously broken axial symmetry of the vortex core -- the twist oscillations of the vortex core propagating along the vortex line. This Figure is also applicable to the ${\cal N}=2$ vortex in $^3$He-B , which consists of two composite objects -- spin-mass vortices connected by the topological soliton,\cite{Kondo1992}  see Fig. \ref{TopObjectsB}. 
(g) Half-quantum vortices  in the polar phase of $^3$He, which  in a tilted magnetic field are connected by the topological spin soliton.\cite{Autti2016} Red arrows show direction of magnetic field, and blue arrows -- direction of spin-nematic vector $\hat{\bf d}$, which rotates by $\pi$ around each half-quantum vortex.
 }
 \label{TopObjects}
\end{figure}

The classification of the topological objects in the order parameter fields  revealed the possibility of many configurations with nontrivial topology, which are described by the homotopy groups\cite{1976b,1977a,1977b} and by the relative homotopy groups \cite{1978a,Volovik1978}. Some of the topological defects and topological textures are shown in Fig. \ref{TopObjects}.

All three superfluid phases discussed here are the spin-triplet $p$-wave
superfluids, i.e. the Cooper pair has spin $S=1$ and orbital momentum $L=1$.
The order parameter is given by $3\times 3$ matrix $A_{\alpha i}$ (see Eq.(\ref{Triplet})), which transforms as a vector under $SO(3)_S$ spin rotations (first index) and as a vector under $SO(3)_L$ orbital rotations (second index).
The order parameter $A_{\alpha i}$ comes as the bilinear combination of the fermionic opertators, $A_{\alpha i} \propto \left<\psi \sigma_\alpha \nabla_i \psi\right>$, where $\sigma_\alpha$ are Pauli matrices for the nuclear spin of $^3$He atom. Note that similar order parameter appears in the so-called spinor quantum gravity, where it describes the emergent tetrad field.\cite{Akama1978,Volovik1990b,Wetterich2005,Diakonov2012,Wetterich2013}

\subsection{Chiral superfluid $^3$He-A}

In the ground state of $^3$He-A the order parameter matrix has the form
\begin{equation}
A_{\alpha i}= \Delta_A e^{i\Phi} \hat d_\alpha(\hat e_1^i + i \hat e_2^i)~~,~~ \hat{\bf l}=\hat{\bf e}_1\times \hat{\bf e}_2\,,
\label{Aphase}
\end{equation}
where $\hat{\bf d}$ is the unit vector of the anisotropy in the spin space due to  spontaneous breaking of $SO(3)_S$ symmetry; $\hat{\bf e}_1$ and $\hat{\bf e}_2$ are mutually orthogonal unit vectors;    and $\hat{\bf l}$ is the unit vector of the anisotropy in the orbital space due to  spontaneous breaking of $SO(3)_L$ symmetry. The $\hat{\bf l}$-vector also shows the direction of the orbital angular momentum of the chiral superfluid, which emerges due to spontaneous breaking of time reversal symmetry.  
The chirality of $^3$He-A has been probed in several experiments.\cite{Walmsley2012,Kono2013,Kono2015}

In the chiral superfluid the superfluid velocity $ {\bf v}_{\rm s}$ of the chiral condensate is determined not only by the condensate phase $\Phi$, but also by  the  orbital triad  $\hat{\bf e}_1$, $\hat{\bf e}_2$ and $\hat{\bf l}$:
\begin{equation}
  {\bf v}_{\rm s}=\frac{\hbar}{2m}\left(\nabla\Phi+\hat{\bf e}_1^i\nabla \hat{\bf e}_2^i \right)\,,
\label{SuperfluidVelocity}
\end{equation}
where $m$ is the mass of the $^3$He atom. 
As distinct form the non-chiral superfluids, where the vorticity is presented in terms of the quantized singular vortices with the phase winding $\Delta\Phi = 2\pi {\cal N}$ around the vortex core, in $^3$He-A the vorticity can be continuous. The continuous vorticity is represented by the texture of the unit vector $\hat{\bf l}$ according to the Mermi-Ho relation:\cite{Mermin-Ho}
\begin{equation}
  \nabla\times{\bf v}_{\rm s} =
     \frac{\hbar}{4m}e_{ijk} \hat l_i{\bf \nabla} \hat l_j\times{\bf
\nabla}
\hat l_k \,.
\label{Mermin-HoEq}
\end{equation}

\begin{figure}
 \includegraphics[width=0.5\textwidth]{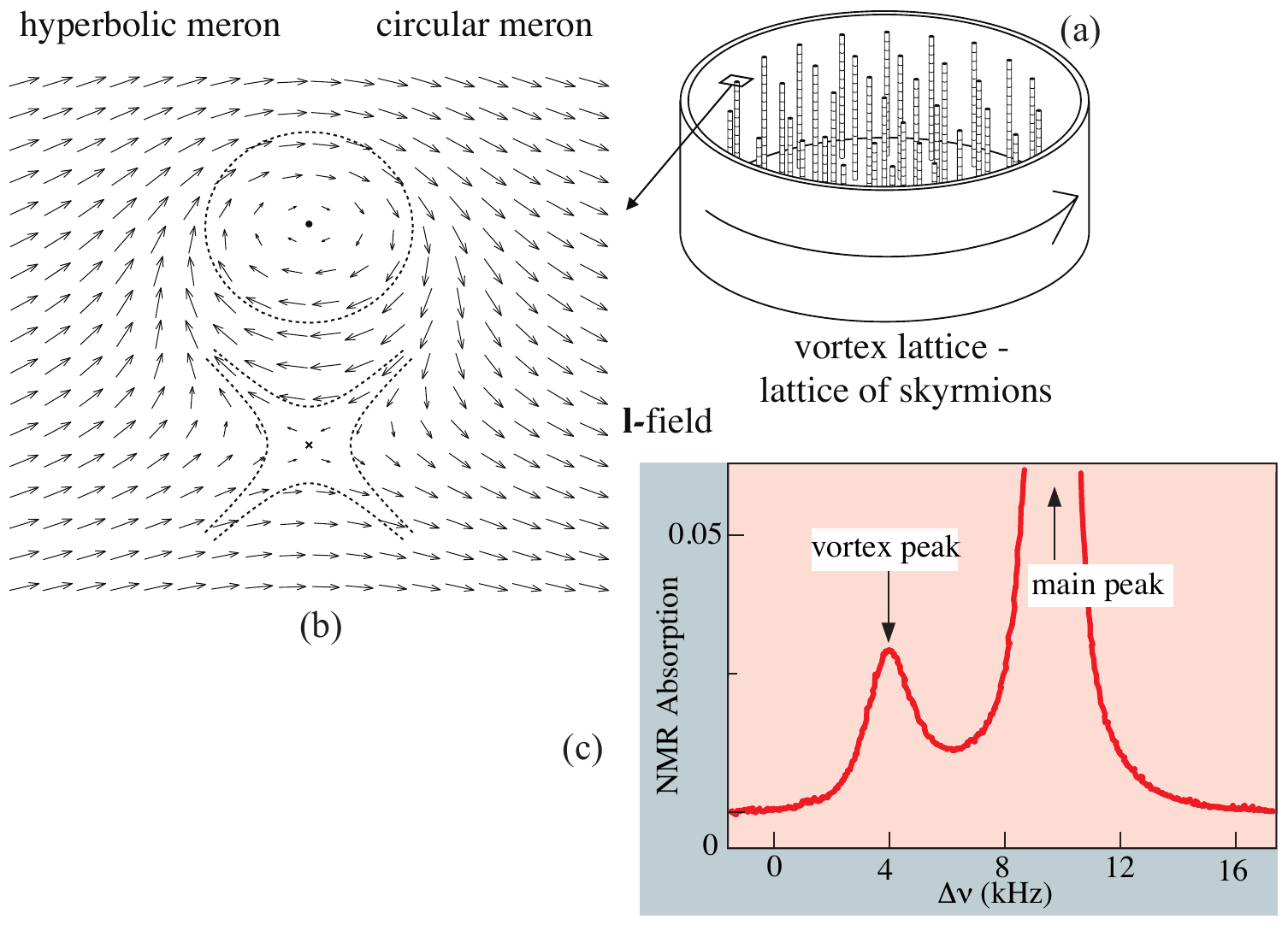}
 \caption{
(a) Rotating state in $^3$He-A in high magnetic field as an array of isolated skyrmions with $m_l=1$ and ${\cal N}=2$ in Fig. \ref{TopObjects}a. (b) Such skyrmion contains two merons -- circular and hyperbolic.
(c) NMR signature of vortex skyrmion -- the satlellite peak in the NMR spectrum.
The type of the skyrmion is identified by the position of the satellite peak, the peak amplitude reflects the number of skyrmions in the sample.
In the absence of magnetic field the skyrmions merge forming the periodic texture in Fig. \ref{FourMerons} with 
 ${\cal N}=4$ in the elementary cell.
 }
 \label{Skyrmion2merons}
\end{figure}

Vorticity is created in rotating cryostat. In $^3$He-A, the continuous textures are more easily created than the singular objects with the hard core of the coherence length size $\xi$, which formation requires overcoming of large energy barrier. That is why  the typical objects which appear under rotation of cryostat with $^3$He-A  is the vortex-skyrmion. It is the continuous texture of the orbital $\hat{\bf l}$-vector in Fig. \ref{TopObjects}a without any singularity in the order parameter fields. This texture represents the  vortex with doubly quantized (${\cal N}=2$) circulation of superfluid velocity around the texture,
$\oint d{\bf r}\cdot {\bf v}_{\rm s} ={\cal N}\kappa$, where $\kappa=h/2m$ is the quantum of circulation.\cite{Chechetkin1976,AndersonToulouse1977} 
The vortex-skyrmions have been identified in rotating cryostat in 1983.\cite{Seppala1983,Seppala1984}
They are described by two topological invariants, in terms of the orbital vector  $\hat{\bf l}$ and in terms of the spin nematic vector $\hat{\bf d}$:
\begin{equation}
m_l= \frac{1}{4\pi}
\int dx~dy~
\hat{\bf l}\cdot\left(\frac{\partial\hat{\bf l}}{\partial x}  \times
\frac{\partial\hat{\bf l}}{\partial y} \right)= \frac{1}{2}{\cal N}\,,
\label{ltextureinvariant}
\end{equation}
\begin{equation}
m_d= \frac{1}{4\pi}
\int dx~dy~
\hat{\bf d}\cdot\left(\frac{\partial\hat{\bf d}}{\partial x}  \times
\frac{\partial\hat{\bf d}}{\partial y} \right)\,.
\label{dtextureinvariant}
\end{equation}
The last equality in Eq.(\ref{ltextureinvariant}) shows the connection between the topological charge of the orbital texture and the circulation of superfluid velocity around it, which follows from the Mermin-Ho relation (\ref{Mermin-HoEq}).

\begin{figure}
 \includegraphics[width=0.5\textwidth]{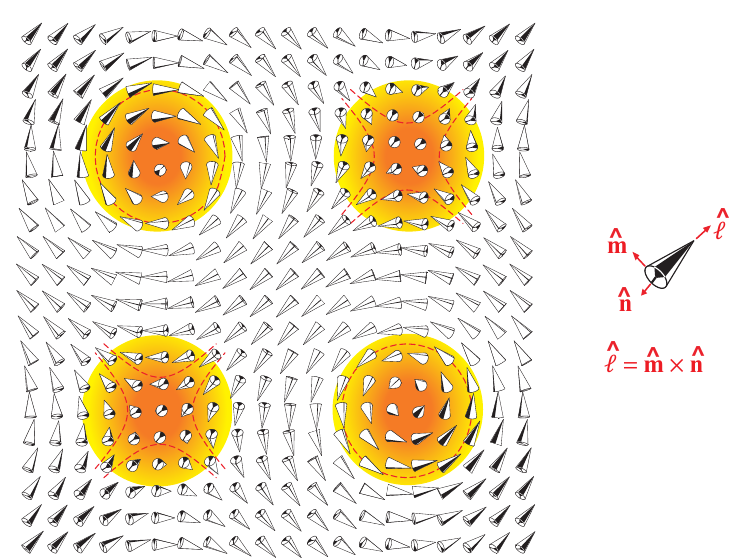}
 \caption{Elementary cell of meron lattice in rotating superfluid $^3$He-A in the absence of magnetic field (adapted from \cite{Eltsov2000}). The unit cell  consists of four merons,  two of which are circular and two are hyperbolic. In each meron the orbital vector $\hat{\bf l}$  covers half a sphere, and according to
 Eq.(\ref{Mermin-HoEq}) each meron represents a vortex with a single quantum of circulation, 
${\cal N}=1$. Thus the unit cell carries topological charge $m_l=2$ and thus the circulation number ${\cal N}=4$.   In the circular vortex-meron the orbital vector $\hat{\bf l} \parallel \mathbf{\Omega}$ in the center, where $\mathbf{\Omega}$
is the angular velocity of the rotating cryostat.  In the hyperbolic vortex-meron $\hat{\bf l} \parallel - \mathbf{\Omega}$ in the center. 
When the magnetic field is switched on, this state transforms to an array of isolated vortex-skyrmions with ${\cal N}=2$ each in  Fig. \ref{TopObjects}a. Such skyrmion contains two merons, see  Fig. \ref{Skyrmion2merons}b.
 }
 \label{FourMerons}
\end{figure}

In a high magnetic field the vortex lattice consists of isolated vortex-skyrmions with ${\cal N}=2$, $m_l=1$ and $m_d=0$, see Fig. \ref{Skyrmion2merons}. In the low field, when the magnetic energy is smaller than the spin-orbit interaction, the vortex-skyrmion  with $m_d=m_l=1$ becomes more preferable. 
The first order topological transition, at which the topological charge $m_d$ of the skyrmion changes from 0 to 1, has been observed in acoustic experiments.\cite{Pekola1990}
Finally, when magnetic field is close to zero, the skyrmions are not isolated. They form the periodic 
vortex texture represented in terms of merons -- the continuous Mermin-Ho vortices with 
$m_l=1/2$ and ${\cal N}=1$ each. The elementary cell of the vortex structure of rotating $^3$He-A contains four merons, see Fig. \ref{FourMerons}. It has topological charge $m_l=m_d=2$ and ${\cal N}=4$.
The isolated skyrmion in the non-zero field in Fig. \ref{TopObjects}a can be represented as the bound state of two merons in Fig.\ref{Skyrmion2merons}.

In 1994  new type of continuous vorticity has been observed  in $^3$He-A -- the vortex texture in the form of the vortex sheets, Fig. \ref{TopObjects}c.\cite{Parts1994a,Parts1994b}  Vortex sheet is the topological soliton with kinks, each kink representing  the continuous Mermin-Ho vortex with ${\cal N}=1$, Fig. \ref{TopObjects}d.

In addition to continuous vortex textures, the rotating state of $^3$He-A may consist of the singular vortices with ${\cal N}=1$.
They are observed in NMR experiments if one starts rotation in the normal phase and then cools down to the A-phase.\cite{Parts1995}
In principle the same scheme can lead to the formation of the half-quantum vortices (HQVs), which have been suggested to exist  in thin films of $^3$He-A.\cite{1976b}  
The half-quantum vortex represents the condensed matter analog of the Alice string in particle physics.\cite{Schwarz1982}  The half-quantum vortex is the vortex with fractional circulation of superfluid velocity, ${\cal N}=1/2$. It is topologically confined with the fractional spin vortex, in which $\hat{\bf d}$ changes sign when circling around the vortex:
\begin{equation}
\hat{\bf d}(\hat{\bf r}) e^{i\Phi(\hat{\bf r})}=\left(\hat{\bf x} \cos\frac{\phi}{2}+ \hat{\bf y} \sin\frac{\phi}{2}\right)e^{i\phi}
\,,
\label{HalfQuantumVortex}
\end{equation}
When the azimuthal coordinate $\phi$ changes from 0 to $2\pi$ along the circle around this object, the vector $\hat{\bf d}(\hat{\bf r})$  changes sign and simultaneously the phase $\Phi$ changes by $\pi$, giving rise to ${\cal N}=1/2$. The  order parameter (\ref{HalfQuantumVortex}) remains continuous along the circle.
While a particle that moves around an Alice string flips its charge, the quasiparticle moving around the half-quantum vortex flips its spin quantum number. This gives rise to the Aharnov-Bohm effect for spin waves
in NMR experiments.\cite{SalomaaVolovik1987} 

In superfluid $^3$He the HQVs have been stabilized  only recently and in a different phase -- in the polar phase of $^3$He confined in aerogel \cite{Autti2016}, see Sec.\ref{PolarVortices}. HQVs have been identified in NMR experiments due to the topological soliton, which is attached to the spin vortex in tilted magnetic field  because of spin-orbit interaction, Fig. \ref{TopObjects}g.

Another object which is waiting for its observation in $^3$He-A is the vortex terminated by hedgehog.\cite{Blaha1976,1976a} This is the condensed matter analog of the electroweak magnetic monopole and the other monopoles connected by strings.\cite{Kibble2008} The hedgehog-monopole, which terminates the vortex, exists in particular at the interface between $^3$He-A and $^3$He-B, Fig. \ref{TopObjects}f. The topological defects living on the surface of the condensed matter system or at the interfaces are called boojums.
\cite{Mermin1977} They are classified in terms of relative homotopy groups.\cite{Volovik1978}   Boojums in Fig. \ref{TopObjects}f terminate the 
$^3$He-B vortex-strings with ${\cal N}=1$. Though boojums do certainly exist on the surface of rotating $^3$He-A and at the interface between the rotating $^3$He-A and $^3$He-B, at the moment their NMR signatures are too weak to be resolved in NMR experiments. Experimentally the vortex terminated by the hedgehog-monopole was observed in cold gases.\cite{Mottonen2014}

\begin{figure}
 \includegraphics[width=0.5\textwidth]{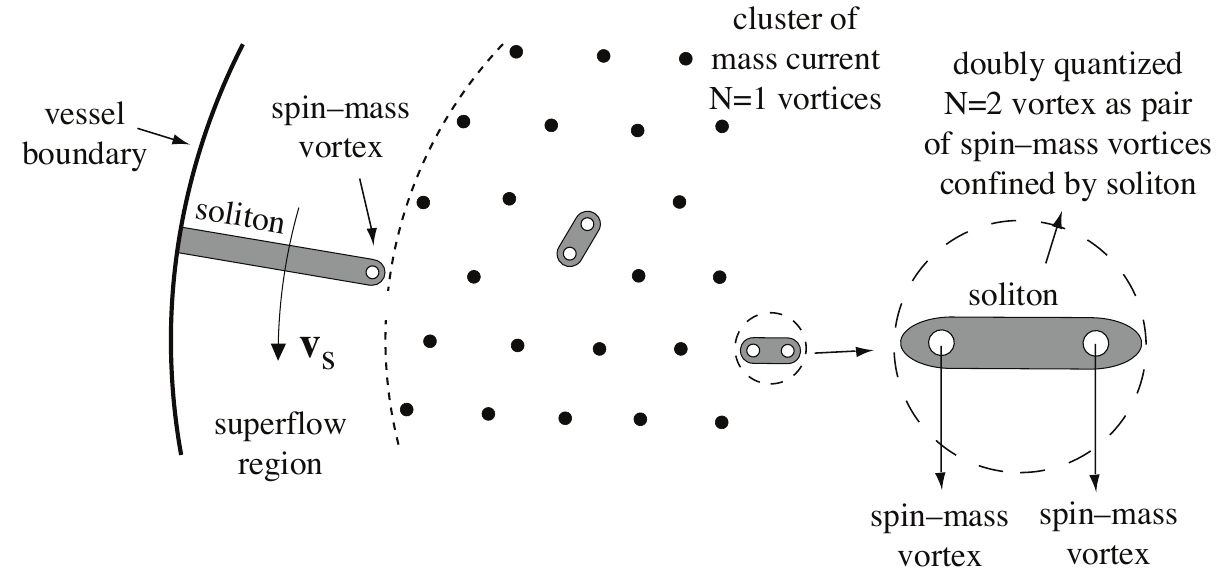}
 \caption{Topological oblects observed in superfluid  $^3$He-B in rotating cryostat.
 Conventional mass current vortices with ${\cal N}=1$ form a regular
structure. If the number of vortices is less than
equilibrium number for a given rotation velocity, vortices are collected  in
the vortex cluster with the vortex free region outside the cluster, where the mass current is circulating.  
Spin vortices, which have the soliton tail, are stabilized being pinned by the cores of the mass current vortices.
They form the composite object -- the spin-mass vortex with the soliton tail.\cite{Kondo1992}
 A single spin--mass vortex is stabilized at
the periphery of the cluster by the combined effect of soliton tension
and the Magnus force acting on the mass vortex from the superflow in the vortex-free region. 
Pair of spin-mass vortices connected by soliton forms the doubly quantized ${\cal N}=2$
mass vortex.
 }
 \label{TopObjectsB}
\end{figure}

\subsection{Superfluid $^3$He-B}

In the ground state of $^3$He-B the order parameter matrix has the form
\begin{equation}
A_{\alpha i}= \Delta_B e^{i\Phi} R_{\alpha i}\,,
\label{Bphase}
\end{equation}
where $R_{\alpha i}$ is the real matrix of rotation, $R_{\alpha i}R_{\alpha j}=\delta_{ij}$.

Vorticity  in the non-chiral superfluid is always singular, but in $^3$He-B  it is also presented in several forms. Even the ${\cal N}=1$ vortex, where $\Phi(\hat{\bf r})=\phi$, has an unusual structure of the singular vortex core. Already in the first experiments with rotating 
 $^3$He-B the first order phase transition has been observed, which has been associated with the transition inside the vortex core.\cite{Ikkala1982}
It was suggested that at the transition the vortex core becomes 
non-axisymmetric, i.e. the axial symmetry of the vortex is spontaneously broken in the vortex core.\cite{Thuneberg1986,VolovikSalomaa1985} This was confirmed in the further experiments, where the Goldstone mode associated with the symmetry breaking  was identified -- the twist oscillations in  Fig. \ref{TopObjects}f and Fig. \ref{Witten} propagating along the vortex line.\cite{Kondo1991} 
In the weak coupling BCS theory, which is applicable at low pressure, such vortex can be considered as splitted into two half-quantum vortices connected by non-topological soliton.\cite{Volovik1990,SilaevThuneberg2015}

On the  other side of the transition,  at high pressure,  the structure of the vortex core in $^3$He-B is axisymmetric, but it is also nontrivial. In the core the discrete symmetry is spontaneously broken, as a result the core does not contain the normal liquid, but is occupied by the chiral superfluid -- the A-phase of $^3$He.\cite{SalomaaVolovik1987} The A-phase core in axisymmetric vortex results in the observed large magnetization of the  core compared with that in the non-axisymmetric vortex.\cite{MagneticVortices1983}
Later we shall discuss the Weyl fermions living within such a core in Sec. \ref{AtoB} and Sec. \ref{multple}.

\begin{figure}
 \includegraphics[width=0.5\textwidth]{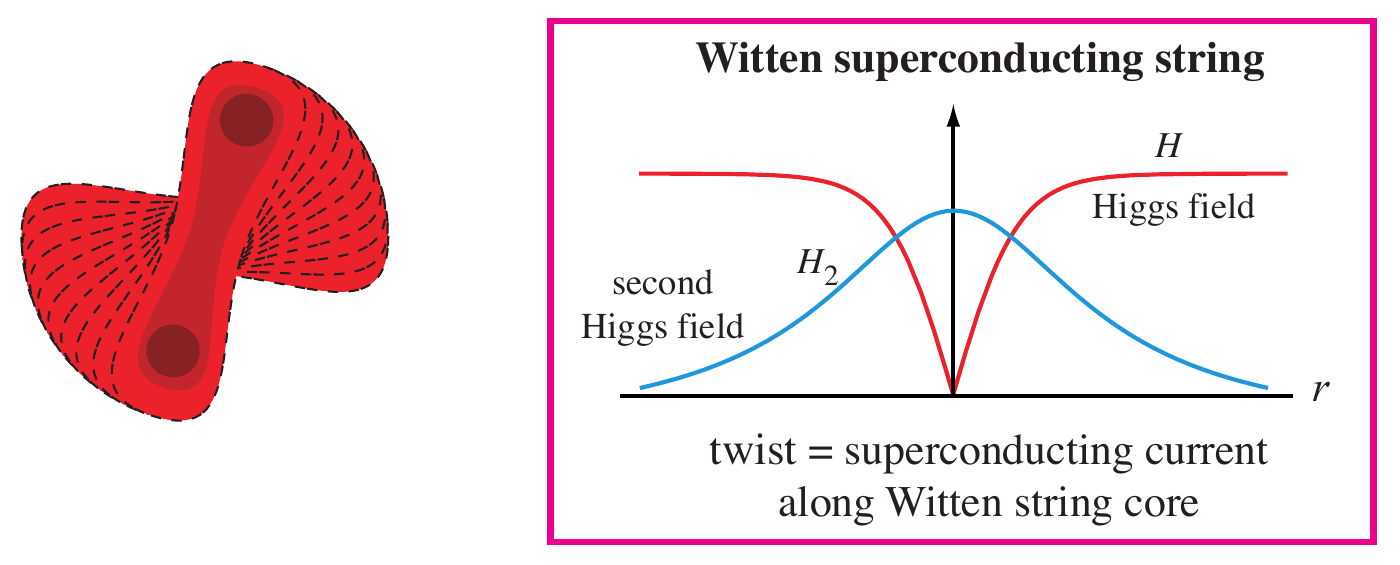}
 \caption{Twist of the asymmetric vortex core with sponaneously broken $SO(2)$ symmetry observed in $^3$He-B\cite{Kondo1991}  is analogous to the superconducting electric current along
the Witten cosmic string with spontaneously broken $U(1)$ symmetry. The second Higgs field  $H_2=|H_2|e^{i\Phi_2}$, which
spontaneously appears in the core of the Witten string, breaks the electromagnetic $U(1)$ symmetry giving rise to the 1D superconductivity  with the current ${\bf j}\propto \hat{\bf z}\,\partial_z\Phi_2$ concentrated in the string core.\cite{Witten1985} 
 }
 \label{Witten}
\end{figure}

The phenomenon of the additional symmetry breaking in the core of the topological defect has been also discussed for cosmic strings.\cite{Witten1985} The spontaneous breaking of the electromagnetic $U(1)$ symmetry in the core of the cosmic string has been considered, due to which the core becomes superconducting with supercurrent along the core. The string with the superconducting electric current is analogous to the asymmetric vortex with twisted core, see Fig. \ref{Witten}.

The topology of $^3$He-B also admits existence of spin vortices, $Z_2$ topological defects of the matrix $R_{\alpha i}$. Due to spin-orbit interaction, which violates the invariance under $SO(3)_S$ spin rotations, the spin vortex gives rise to the topological soliton attached to the vortex line, similar to that in  Fig. \ref{TopObjects} for the polar phase. That is why, if the spin vortex appears in the cell, it is pushed to the wall of the container by the soliton tension and disappears at the wall. However, the spin vortex survives if it is pinned by the conventional vortex (mass current vortices) with ${\cal N}=1$ and forms the composite object -- the spin-mass vortex.
Experimentally two types of composite objects have been identified in  $^3$He-B, see Fig. \ref{TopObjectsB}.
(i) The other end of the soliton is at the wall of container. (ii) The ${\cal N}=2$ vortex is formed which consists of two spin-mass vortices connected by soliton.\cite{Kondo1992} 

\subsection{Polar phase of superfluid $^3$He}
\label{PolarVortices}

Polar phase of superfluid $^3$He has been  stabilized in a nematically ordered aerogel with nearly parallel strands (nafen).\cite{HalperinParpiaSauls2018,Halperin2019,Dmitriev2015} 
In the ground state of the polar phase the order parameter matrix has the form
\begin{equation}
A_{\alpha i}= \Delta_P e^{i\Phi} \hat d_\alpha \hat z^i \,,
\label{PolarPhase}
\end{equation}
where axis $z$ is along the nafen strands.
Topology of polar phase in nafen suggests existence of ${\cal N}=1$ mass current vortices, $Z_2$ spin vortices and the half-quantum vortices. The latter is the combination of the fractional ${\cal N}=1/2$ mass vortex and the fractional spin vortex in Eq.(\ref{HalfQuantumVortex}). The spin-orbit interaction in the polar phase is more preferable for the half-quantum vortices than in  $^3$He-A.  In the absence of magnetic field, or if the field is along the nafen strands the
spin-orbit interaction does not lead to formation of the solitons attached to the spin vortices. As a result the half-quantum vortices become emergetically favorable and appear in the rotating cryostat if the sample is cooled down from the normal state under rotation. The HQVs are identified due to peculiar dependence of the NMR frequency shift on the tilting angle of magnetic field.\cite{Autti2015}
Fig. \ref{TopObjects}g shows a pair of half-quantum vortices in transverse magnetic field (red arrows). Blue arrows show the distribution of the nematic vector $\hat{\bf d}$ of the spin part of the order parameter in the polar phase.

\begin{figure}[t]
\includegraphics[width=\linewidth]{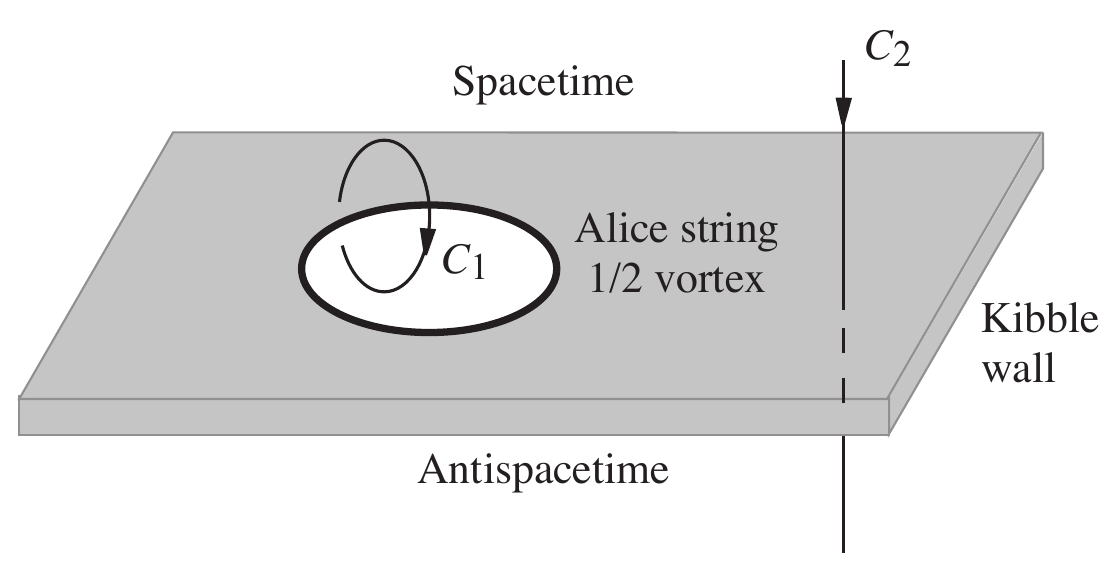}
\caption{  In $^3$He-B,  the half-quantum vortex (analog of Alice string) looses its topological stability and becomes the termination line of a non-topological domain wall -- the Kibble wall \cite{Makinen2018}. In terms of the tetrads, the Kibble wall separates the states with different tetrad determinant, and thus between the "spacetime" and "antispacetime" \cite{Volovik2019}. There are two roads to antispacetime: the "safe" route around the Alice string (along the contour $C_1$) or "dangerous" route along $C_2$ across the Kibble wall.
}
\label{TwoRoads_Fig}
\end{figure}

Later it was found \cite{Makinen2018} that the half-quantum vorices survive the phase transition to $^3$He-B, where the half-quantum vortex is topologically unstable. In the B-phase the half-quantum vortices pinned by the starnds of nafen become the termination lines of the non-topological domain walls - the analog of Kibble cosmic walls\cite{Kibble1982}. In $^3$He-B, the Kibble wall separates the states with different tetrad determinant, and thus between the "spacetime" and "antispacetime" \cite{Volovik2019}. More on spacetime in cosmology and condensed matter see
\cite{Turok2018,Turok2018b,Rovelli2012b,Rovelli2012a,NissinenVolovik2018,Vergeles2019,Volovik2019b}.

\section{Momentum space topology}

The topological stability of the coordinate dependent objects -- defects and textures -- is determined by the pattern of the symmetry breaking in these superfluids. Now we shall discuss these three phases of superfluid $^3$He from the point of view of momentum-space topology, which desribes the topological properties of the homogeneous ground state of the superfluids. These superfluids represent three types of topological materials, with different geometries of the topologically protected nodes in the spectrum of fermionic quasiparticles: Weyl points in $^3$He-A, Dirac lines in the polar phase and Majorana nodes on the surface of $^3$He-B.

\begin{figure}
\centerline{\includegraphics[width=1.0\linewidth]{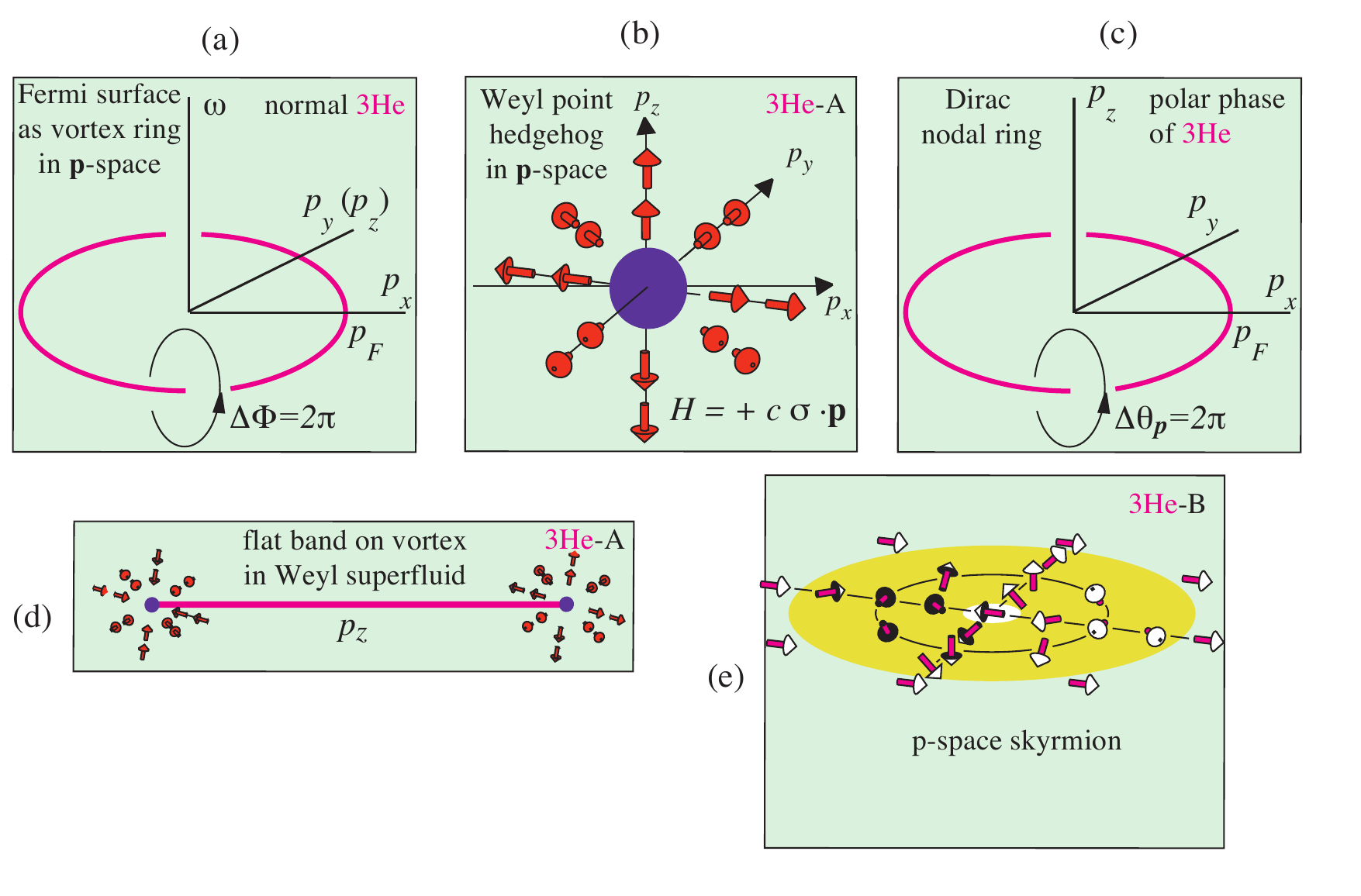}}
\caption{ 
\label{Fig:classes} 
 Topological materials as configurations in momentum space.  
(a)  Fermi surface in normal liquid $^3$He is topologically protected, since
the Green's function has singularity in the form of the  vortex ring in the $({\bf p},\omega)$-space.
(b)  Weyl point in $^3$He-A as the hedgehog in ${\bf p}$-space. It can  be also reperesented as the  ${\bf p}$-space Dirac monopole with the Berry 
magnetic flux.
(c) The polar phase has the Dirac nodal line in ${\bf p}$-space -- the counterpart of the spin vortex in real space.
(d) The ${\cal N=1}$ singular vortex in chiral superfluid $^3$He-A  has the one-dimensional flat band terminated by the projections of the Weyl points
to the vortex line.\cite{KopninSalomaa1991,Volovik1994,Volovik2011}  In real space it has analogy with the Dirac monopole terminating the Dirac string.
(e) Skyrmion configurations in ${\bf p}$-space describe  the fully gapped topological superfluids.  The 2D skyrmion  describes the topology of the $^3$He-A state in a thin film and
the topology of 2D planar phase of $^3$He.\cite{VolovikYakovenko1989,Volovik1989,Yakovenko1989} 
 The 3D ${\bf p}$-space skyrmion describes the superluid 
 $^3$He-B.\cite{SalomaaVolovik1988,Schnyder2008,Kitaev2009,Mizushima2015}  
}
\end{figure}

The properties of the fermionic spectrum in the bulk or/and on the surface of superfluids is determined by the topological properties of the Bogoliubov-de Gennes Hamiltonian
\begin{equation}
H({\bf p})=  
\begin{pmatrix} 
\epsilon({\bf p}) & \hat\Delta({\bf p})
\\ 
\hat\Delta^+({\bf p}) & -\epsilon({\bf p})
\end{pmatrix}  \,, \,\epsilon({\bf p})=\frac{p^2 -p_F^2}{2m^*} \,,
\label{H}
\end{equation}
or by the Green's function
 \begin{equation}
G^{-1}(\omega,{\bf p})=i\omega -H
\,.
\label{eq:GreenFunction}
\end{equation}
For the spin triplet $p$-wave superfluid $^3$He the gap function is expressed in terms of the $3\times 3$ order parameter matrix $A_{\alpha i}$:
\begin{equation}
\hat\Delta({\bf p})=A_{\alpha i}  \sigma_\alpha  \frac{p_i}{p_F} ~.
\label{Triplet}
\end{equation}

The topologically stable singularities of the Hamiltonian or of the Green's function in the momentum or momentum-frequency spaces look similar to the real-space topology of the defects and textures, see Fig. \ref{Fig:classes}. 
The Fermi surface, which describes the normal liquid $^3$He and metals, represents the topologically stable 2D object in the 4D frequency-momentum space $(p_x,p_y,p_z,\omega)$, is analogous to the vortex ring. The Weyl points in $^3$He-A 
 represents the topologically stable hedgehog in the 3D ${\bf p}$-space, Fig. \ref{Fig:classes}b.  
The Dirac nodal line in the polar phase of $^3$He is the ${\bf p}$-space analog of the spin vortex, Fig. \ref{Fig:classes}c. The nodeless 2D systems (thin films of $^3$He-A and the planar phase) and the nodeless 3D system -- $^3$He-B -- are characterized by the topologically nontrivial skyrmions in momentum space
in Fig. \ref{Fig:classes}e.

\subsection{Normal liquid $^3$He}

The normal state of liquid $^3$He belongs to the class of Fermi liquids, which properites at low energy are determined by quasiparticles living in the vicinity of the Fermi suraface. The systems with Fermi surface, such as metals,  are the most widespread topological materials in nature. The reason for that is that the Fermi surface  is topologically protected and thus is robust to small perturbations. This can be seen on the simple example of the 
Green's function for the Fermi gas at the imaginary frequency:
\begin{equation}
 G^{-1}(\omega,{\bf p})=i\omega - \left(\frac{p^2}{2m} -\mu\right)  \,.
\label{GreenFermiGas}
\end{equation}
The Fermi surface at $p=p_F$ exists at positive chemical potential, with $p^2_F /2m=\mu$.
The topological protection is demonstrated in 
Fig.\ref{Fig:classes}a for the case of the 2D Fermi gas, where the Fermi surface is the line $p=p_F$  in $(p_x,p_y)$-space.
 In the extended $(\omega,p_x,p_y)$-space this gives rise to singularity in the Green's function on the line at which $\omega=0$ and $p=p_F$, where the Green's function is not determined. Such singular  line in momentum-frequency space looks similar to the vortex line in real space: the phase 
$\Phi({\bf p},\omega)$ of the Green's function $G({\bf p},\omega)=|G({\bf p},\omega)| e^{i\Phi({\bf p},\omega)}$ changes by $2\pi$ around this line. 
In general, when the Green's function is the matrix with spin or/and band indices, the integer valued topological invariant -- the winding number of the Fermi surface --
has the following form:\cite{Volovik2003}
\begin{equation}
N={\bf tr}~\oint_C {dl\over 2\pi i}  G(\omega,{\bf p})\partial_l
G^{-1}(\omega,{\bf p})~.
\label{InvariantForFS}
\end{equation}
Here the integral is taken over an arbitrary contour $C$ around the Green's function singularity
in  the $D+1$
momentum-frequency space.
Due to nontrivial topological invariant, Fermi surface survives the perturbative interaction and exists in the Fermi liquid as well. Moreover the singularity in the Green's function remains if due to interaction  the Green's function has no poles, and thus quasiparticles are not well defined. The systems without poles include the marginal Fermi liquid, Luttinger liquid, and the Mott pesudogap state.\cite{KuchinskiiSadovskii2006}

\begin{figure}
\includegraphics[width=1.0\linewidth]{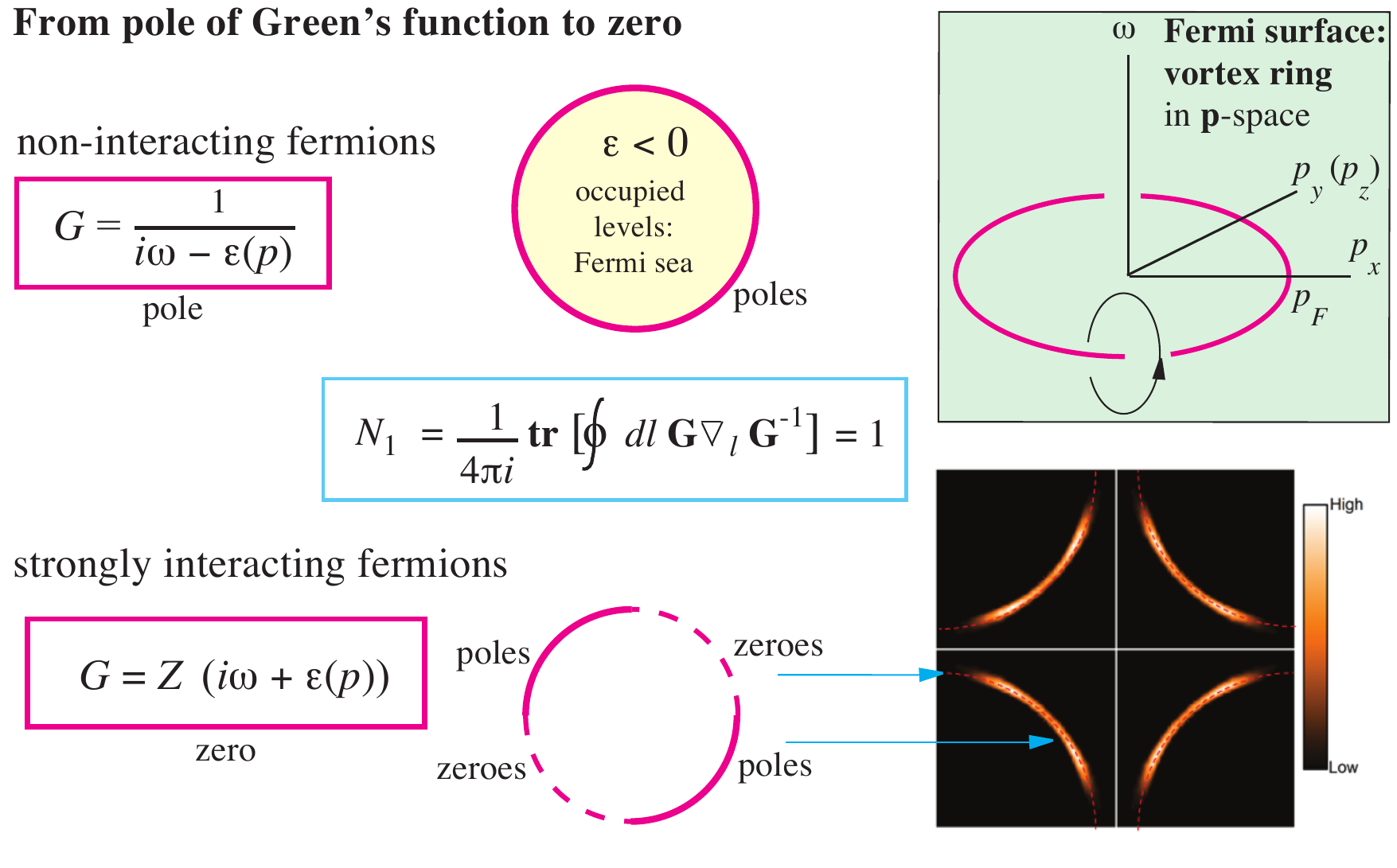}
\caption{Due to interaction, in some parts of the Fermi surface, the poles in the Green's function may transform to zeroes. The topological charge of these parts of Fermi surface remain the same, but they become invisible, while the other parts of the Fermi surface look as Fermi arcs.
 }
 \label{FermiArc}
\end{figure}

It is possible that in the Mott pesudogap state, the poles of the Green's function transform to zeroes of  the Green's function. The topological invariant remains the same, which is the reason why the Luttinger theorem is still valid.\cite{Dzyaloshinskii2003,Farid2009}
The particle density of interacting
fermions is equal to the volume in the momentum
space enclosed by the singular surface with the topological charge $N=1$, irrespective of the realization of the singularity. As distinct from the pole, the zero in the Green's function is invisible, so that the pole region of the Fermi surface looks as the Fermi arc, see Fig.\ref{FermiArc}({\it bottom}).

The Fermi surface may disappear in the topological  quantum phase transition, when the chemical potential $\mu$ crosses zero: the vortex ring object shrinks to the point at $\mu=0$ and disappears at $\mu<0$ if the point is not protected by another topological invariant discussed in Sec. \ref{WeylSuperfluid}.
The Fermi surface may disappear also due to non-perturbative process of the symmetry breaking phase transition, when the fermionic spectrum of the system is drastically resonstructed, as it happens under the  transition from the normal to the superfluid state.

\subsection{Weyl superfluid $^3$He-A}
\label{WeylSuperfluid}

Under the superfluid transition the $2\times 2$ matrix of the normal liquid Green's function with spin indices
transforms to the $4\times 4$ Gor'kov Green's function.
The simplified Green's function, which describes the topology of the chiral superfluid $^3$He-A, has the 
form:
 \begin{equation}
  G^{-1}=i\omega +   \tau_3\left(\frac{p^2}{2m} -\mu\right) + 
c({\mbox{\boldmath$\sigma$}} \cdot\hat{\bf d})
\left(\tau_1\hat{\bf e}_1\cdot{\bf p}  + \tau_2 \hat{\bf e}_2\cdot{\bf p} \right).
 \label{3He-AGreen}
 \end{equation}
The Pauli matrices $\tau_{1,2,3}$ and $\sigma_{x,y,z}$ correspond to the
Bogoliubov-Nambu spin and ordinary spin of $^3$He atom
respectively;   $\mu=p_F^2/2m$ as before; and  the parameter $c=\Delta_A/p_F$.

Instead of the Fermi surface, now there are two  points in the fermionic spectrum,  at ${\bf K}_\pm=\pm p_F\hat{\bf l}$, where the energy spectrum is nullified and the Green's function is not determined at $\omega=0$.
There are several ways of how to describe the topological protection of these two points.
In terms of the Green's function there is the following topological invariant expressed via integer valued integral over the 3-dimensional surface $\sigma$ around the singular point   in the 4-momentum space   $p_\mu=(\omega,{\bf p})$:\cite{Grinevich1988,Volovik2003}
 \begin{equation}
N = \frac{e_{\alpha\beta\mu\nu}}{24\pi^2}~
{\bf tr}\int_\sigma   dS^\alpha
~ G\partial_{p_\beta} G^{-1}
G\partial_{p_\mu} G^{-1} G\partial_{p_\nu}  G^{-1}\,.
\label{MasslessTopInvariant3D}
\end{equation}
If the invaraint (\ref{MasslessTopInvariant3D}) is nonzero, the Green's function has a singularity inside the surface $\sigma$, and this means that fermions are gapless. The typical singularities
have topological charges $N =+1$ or $N =-1$.
Close to such points the quasiparticles behave as right-handed and left-handed Weyl fermions\cite{Weyl1929} respectively, that is why such point node in the spectrum is called the Weyl point.
The isolated Weyl point  is protected by topological invariant (\ref{MasslessTopInvariant3D}) and survives when the interaction between quasiparticles is taken into account. The Weyl points with the opposite charge $N$ may cancel each other, when they merge together at the quantum phase transition, if some continuous or discrete symmetry does not prohibit the annihilation.

\begin{figure}
\includegraphics[width=1.0\linewidth]{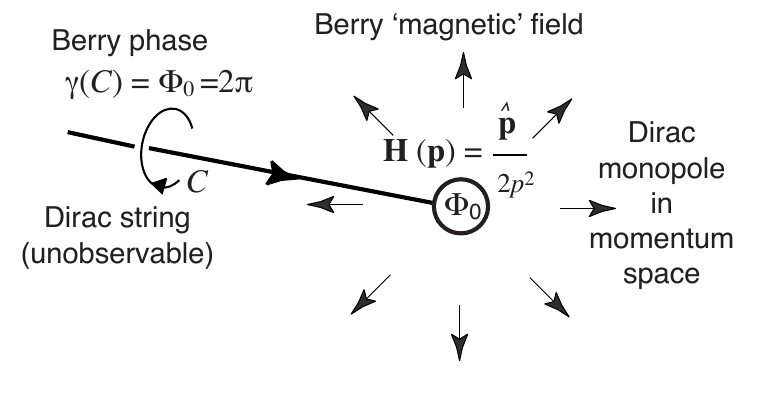}
\caption{Weyl point as magnetic monopole in the Berry phase field.
 }
 \label{BerryMonopole}
\end{figure}
 In $^3$He-A, the topological invariants of the points at ${\bf K}^{(a)}=\pm k_F\hat{\bf l}$ are correspondingly $N =+2$ or $N =-2$: the Weyl points are degenerate over the spin of the $^3$He atoms. 
Considering only single spin projection one comes to the $2\times 2$ Bogoluibov-Nambu Hamiltonian for the spinless fermions:
 \begin{equation}
  H=  {\mbox{\boldmath$\tau$}} \cdot {\bf g}({\bf p}) \,,
 \label{3He-Hamiltonian}
 \end{equation}
where the vector function ${\bf g}({\bf p})$ has the following components in $^3$He-A:
 \begin{equation}
  g_1=\hat{\bf e}_1\cdot{\bf p} ~~,~~g_2= \hat{\bf e}_2\cdot{\bf p} ~~,~~g_3 =\frac{p^2}{2m} -\mu \,.
 \label{3He-Acomponents}
 \end{equation}

The Hamiltonian (\ref{3He-Hamiltonian}) is nullified at two points ${\bf K}^{(a)}=\pm p_F \hat{\bf l}$, where $p=p_F$ and ${\bf p}\cdot \hat{\bf e}_1={\bf p}\cdot \hat{\bf e}_2=0$. At these points the unit vector $\hat{\bf g}({\bf p}) =  {\bf g}({\bf p}) /| {\bf g}({\bf p})|$ has the singularity of the hedgehog-monopole type in Fig. \ref{Fig:classes}b, which
is described by the dimensional reduction of the invariant (\ref{MasslessTopInvariant3D}):
\begin{equation}
N= \frac{1}{8\pi} e_{ikl}
\int_{\sigma} dS^i
\hat{\bf g}\cdot\left(\frac{\partial\hat{\bf g}}{\partial p_k}  \times
\frac{\partial\hat{\bf g}}{\partial p_l} \right)\,,
\label{HedgehogInvariant}
\end{equation}
where $\sigma$ now is the 2D spherical surface around the hedhehog.

The hedgehog has $N=\pm 1$ and it represents the Berry phase magnetic monopole.\cite{Volovik1987,Volovik2003} In the vicinity of the monopole the Hamiltonian can be expanded
in terms of the deviation of the momentum ${\bf p}$ from the Weyl point at ${\bf K}^{(a)}$:\cite{Volovik1986a,Froggatt1991,Horava2005}
\begin{equation}
H^{(a)}=e_\alpha^i \tau^\alpha(p_i-K_i^{(a)})+ \cdots \,.
\label{Atiyah-Bott-Shapiro}
\end{equation}
The emergent linear relativistic spectrum of Weyl fermions leads to the observed $T^4$ behavior of the thermodynamic quantities.\cite{Bartkowiak1999} 
Introducing the effective electromagnetic field ${\bf A}({\bf r},t)=p_F \hat{\bf l}({\bf r},t)$ and effective electric charge $q^{(a)}=\pm 1$, one obtains
\begin{equation}
H^{(a)}=e_\alpha^i \tau^\alpha(p_i-q^{(a)}A_i)+ \cdots \,.
\label{Atiyah-Bott-Shapiro2}
\end{equation}
Such Hamiltonian decsribes the Weyl fermions moving in  the effective electric and magnetic fields
\begin{equation}
{\bf E}_{\rm eff}= -p_F \partial_t\hat{\bf l} ~~,~~{\bf B}_{\rm eff}= p_F \nabla\times\hat{\bf l}\,,
\label{EandB}
\end{equation}  
and also in the effective gravitational field represented by the triad field $e_\alpha^i({\bf r},t)$.
The effective quantum electrodynamics emerging in the vicinity of the Weyl point leads to many analogs in relativistic quantum field theories, including the zero charge effect -- the famous "Moscow zero"  by Abrikosov, Khalatnikov and Landau \cite{LandauAbrikosovKhalatnikov1954}.

In the presence of the superfluid velocity and space-time dependent chemical potential,
the effective spin connection emerges. It enters the long derivative  $D_\mu=\partial_\mu + \frac{1}{4}C_\mu^{\alpha\beta}(\tau_\alpha\tau_\beta - \tau_\beta\tau_\alpha)-iqA_\mu$ with the following non-zero components: $C_i^{12}=m v_{{\rm s}i}({\bf r},t)$ and $C_0^{12}=\mu({\bf r},t)$. The  continuous  vorticity in Eq.(\ref{Mermin-HoEq}) gives rise to the nonzero components of the curvature tensor: $R_{12ik}=m( \partial_iv_{{\rm s}k}- \partial_kv_{{\rm s}i})$.

The Weyl points as the topologically protected touching point of two bands\cite{NeumannWigner1929,Novikov1981} have been discussed in semimetals.\cite{Herring1937,Abrikosov1972,NielsenNinomiya1983,Burkov2011a,Burkov2011b,Weng2015,Huang2015,Lv2015,Xu2015,Lu2015}.
According to the bulk-edge and bulk-defect correspondence, the Weyl points in bulk may produce the Fermi arc on the surface of the material,\cite{Burkov2011a,Burkov2011b} and the flat band of fermionic excitations in the vortex core in Sec. \ref{FlatBandVortex}.

In a different form the topological invariant for the Weyl point has been introduced for the massless neutrino in 1981.\cite{NielsenNinomiya1981} The evidence of neutrino oscillations does not exclude  the possibility that in neutrino sector of particle physics  instead of formation of the Dirac mass there is a splitting of two Weyl points with breaking of CPT symmetry.\cite{KlinkhamerVolovik2005} 

In 1982 the topological invariant for the Weyl points in $^3$He-A have been decsribed in terms of the boojum living on the Fermi surface,\cite{VolovikMineev1982} the analog of the real-space boojum in Fig. \ref{TopObjects}b.
As follows from Eq.(\ref{H}) the node in the spectrum occurs when $\epsilon({\bf  p})=0$ and  the determinant of the gap function is zero, ${\rm det}\,\hat\Delta({\bf p})=0$. That is why the node represents the crossing point of the surface $\epsilon({\bf  p})=0$ (the Fermi surface of the original normal state of liquid $^3$He) and the line  where ${\rm det}\,\hat\Delta({\bf p})=0$. The latter is described by the integer winding number of the  phase $\Phi({\bf p})$ of the determinant:
\begin{equation}
{\rm det}\,\hat\Delta({\bf p})=|{\rm det}\,\hat\Delta({\bf p})| e^{i\Phi({\bf p})} \,,
\label{Determinant}
\end{equation}
 around the nodal line of the determinant.
The topological charge $N$ in Eq.(\ref{MasslessTopInvariant3D}) is nothing but the winding number of the phase $\Phi({\bf p})$  when viewed from the region outside the Fermi surface.\cite{VolovikMineev1982}
Such definition of the topological charge $N$ also coincides with the Berry monopole charge, giving $N= \pm 2$ for the Weyl points at ${\bf K}^{(a)}=\pm p_F\hat{\bf l}$.

In general, the state with the topological charge larger than unity, $|N|>1$, is unstable towards splitting to the states with nodes described elementary charges $N=\pm 1$. In $^3$He-A the nodes with topological charges $N= \pm 2$ are protected by the discrete $Z_2$ symmetry related to spins.

\subsection{Nodal line polar phase}

The topology of the fermionic quasiparticles in the polar phase of $^3$He is described by the following simplified 
Hamiltonian:
\begin{equation}
H=   \tau_3\left(\frac{p^2}{2m} -\mu\right) + 
cp_z({\mbox{\boldmath$\sigma$}} \cdot\hat{\bf d})
\tau_1  \,,
 \label{PolarHamiltonian}
 \end{equation}
where $c=\Delta_P/p_F$.
This Hamiltonian is nullified when $p_z=0$ and $p_x^2+ p_y^2= p_F^2$, i.e. the spectrum of quasiparticles has the nodal line in Fig. \ref{Fig:classes}c. The nodal line is protected by topology due to the discrete symmetry: 
 the   Hamiltonian (\ref{PolarHamiltonian}) anticommutes with $\tau_2$, which allows us to write the  topological charge, see e.g. review \cite{Volovik2013}:
 \begin{equation}
N=  {\bf tr} \oint_C \frac{dl}{4\pi i} ~\tau_2 H^{-1}({\bf p}) \partial_l H({\bf p})\,.
\label{eq:N1}
\end{equation}
Here  $C$ is an infinitesimal contour in momentum space around the line, which is called the Dirac line.
The  topological charge $N$ in Eq.~(\ref{eq:N1}) is integer and is equal to 2 for the nodal line in the polar phase due to spin degeneracy. In a different form the invariant can be found in Ref.\cite{WenZee2002}.

In the moving superfluid, where space and time reversal symmetries are violated, the Dirac ring in the spectrum of Bogoliubov quasiparticles transforms to the Fermi surface -- the so-called Bogoliubov Fermi surface
\cite{Volovik1989e,Agterberg2018}.

Dirac lines exist in cuprate superconductors\cite{Volovik1993} and also in semimetals \cite{HeikkilaVolovik2011,HeikkilaVolovik2016,Mikitik2006,Mikitik2008,Weng2014,Xie2015,Kane2015,Yu2015}.
According to the bulk-edge and bulk-defect correspondence, the Dirac line in bulk may produce the flat band (the band in the fermionic soectrum with exactly zero energy) on the surface of the material\cite{Ryu2002,SchnyderRyu2011,HeikkilaKopninVolovik2011} and the condensation of levels in the vortex core\cite{Volovik2016b}, which we discuss in Sec. \ref{FermionCondensation}. Superconductivity emerging due to the singular density of electronic states in the flat band has been recently found in twisted bi-layer graphene \cite{Cao2018}. The physics of the flat band becomes important for the construction of materials experiencing superconductivity at room temperature \cite{Esquinazi2014,Volovik2018c}.

\subsection{Time reversal invariant fully gapped $^3$He-B}

$^3$He-B belongs to the same topological class as the vacuum of Standard Model in its present insulating phase.\cite{Volovik2010a}  The topological classes of the $^3$He-B states can be represented by the following simplified Bogoliubov - de Gennes Hamiltonian:
 \begin{equation}
H=  \tau_3\left(\frac{p^2}{2m^*} - \mu\right)+  \tau_1
\Delta_B{\mbox{\boldmath$\sigma$}}\cdot\hat{\bf d}({\bf p})\,\, , \,\, \hat d_\alpha({\bf p}) = \pm R_{\alpha i}\frac{p_i}{p_F}
\,.
\label{eq:B-phase}
\end{equation}
Here  $R_{\alpha i}$ is the matrix of rotation; the phase of the order parameter in Eq.(\ref{Bphase}) is chosen either $\Phi=0$ or $\Phi=\pi$;  we also included  the effective mass $m^*$ to discuss the possible topological quantum phase transitions.
In the limit of heavy effective mass, $1/m^* \rightarrow 0$, this model $^3$He-B Hamiltonian transforms to the Hamiltonian for massive relativistic Dirac particles with speed of light  $c_B =\Delta_B/p_F$, and the mass parameter $M=-\mu$.
There are no nodes in the spectrum of fermions: the system is fully gapped. Nevertheless $^3$He-B is the topological superfluid. This can be seen from the integer valued integral over the former Fermi surface:\cite{SalomaaVolovik1988}
\begin{equation}
N_d= \frac{1}{8\pi} e_{ikl}
\int_{S^2} dS^i
\hat{\bf d}\cdot\left(\frac{\partial\hat{\bf d}}{\partial p_k}  \times
\frac{\partial\hat{\bf d}}{\partial p_l} \right)\,,
\label{dinvariant}
\end{equation}
where $N_d=\pm$ depending on the sign in Eq.(\ref{eq:B-phase}).

In terms of the Hamiltonian the topological invariant can be written as integral over the whole momentum space
(or over the Brillouin zone in solids)
\begin{equation}
N_K = {e_{ijk}\over{24\pi^2}} ~
{\bf tr}\left[  \int   d^3p ~K
~H^{-1}\partial_{p_i} H
H^{-1}\partial_{p_j} H H^{-1}\partial_{p_k} H\right],
\label{3DTopInvariant_tau}
\end{equation} 
where $K=\tau_2$ is the matrix which anticommutes with the Hamiltonian. One has $N_K = 2N_d$ due to spin degrees of freedom.
For the interacting systems the Green's function formalism can be used.
In the fully gapped  systems, the Green's function has no singularities in the  whole (3+1)-dimensional space $(\omega,{\bf p})$. That is why we are able to use in Eq.(\ref{3DTopInvariant_tau}) the Green's function at $\omega=0$,
which corresponds to the effective Hamiltonian, $H_{\rm eff}({\bf p})=-G^{-1}(0,{\bf p})$.\cite{Volovik2009a}

In $^3$He-B, the $K=\tau_2$ symmetry   is the combination of time reversal  and particle-hole symmetries. For Standard Model  the corresponding matrix $K$, which anticommutes with the effective Hamiltonian and enters the invariant, is  constructed from the $\gamma$-matrices: it is $K=\gamma_5\gamma^0$.  

 \begin{figure}
 \begin{center}
 \includegraphics[%
  width=1.0\linewidth,
  keepaspectratio]{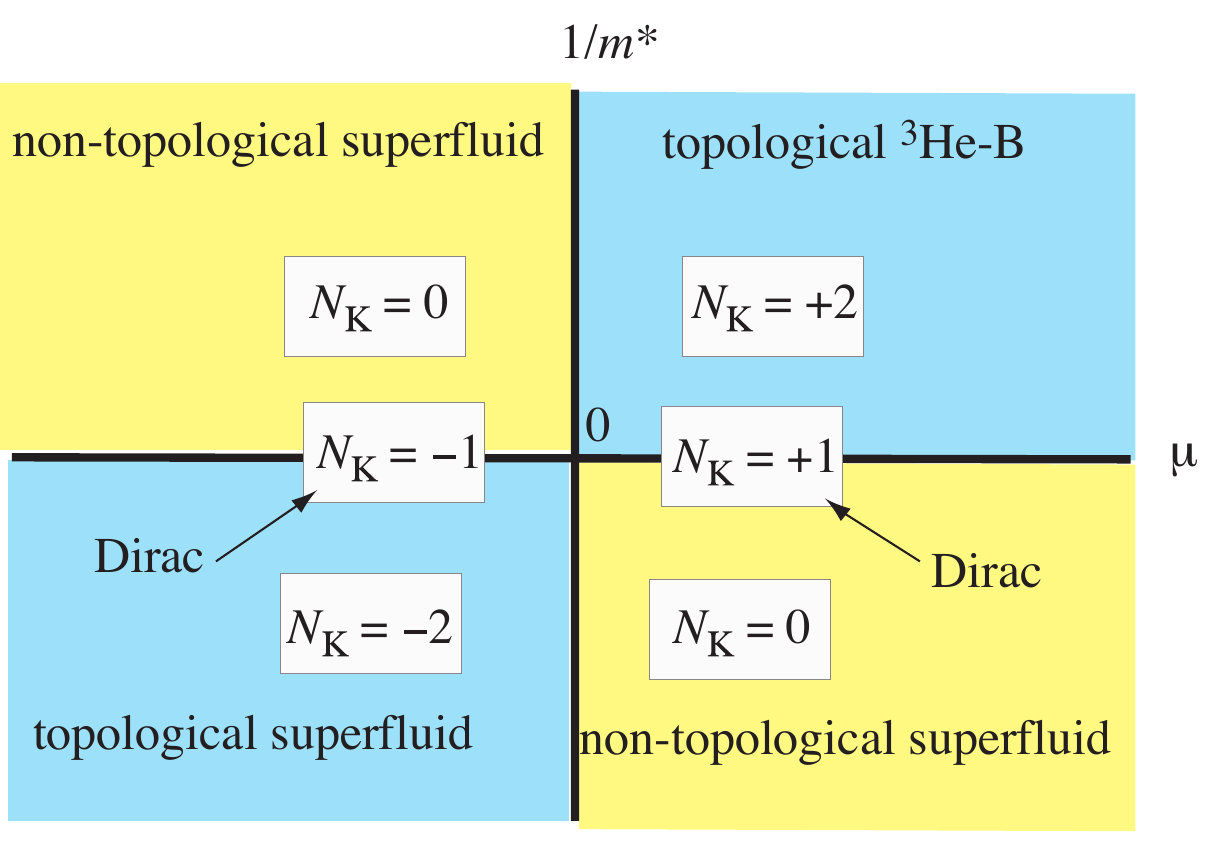}
\end{center}
  \caption{\label{3He-B}  Phase diagram of topological states of $^3$He-B in Eq.(\ref{eq:B-phase}) in the plane $(\mu,1/m^*)$. States on the line
  $1/m^*=0$ correspond to the  Dirac vacua, which Hamiltonian is non-compact. Topological charge of the Dirac fermions
  is intermediate between charges of compact $^3$He-B states.
The line $1/m^*=0$ separates the states with different asymptotic behavior of the Green's function at infinity:
$G^{-1}(\omega=0,{\bf p}) \rightarrow \pm \tau_3 p^2/2m^*$.
 The line $\mu=0$ marks topological quantum phase transition, which occurs between the weak coupling $^3$He-B
 (with $\mu>0$,
 $m^*>0$ and topological charge $N_K=2$) and the strong coupling $^3$He-B   (with $\mu<0$, $m^*>0$ and $N_K=0$).
  This transition is topologically equivalent to quantum phase transition between Dirac vacua with opposite mass parameter
 $M=\pm |\mu|$, which occurs when $\mu$ crosses zero along the line $1/m^*=0$.
 The interface which separates two states contains single Majorana fermion in case of $^3$He-B, and single chiral fermion
 in case of  relativistic quantum fields.  Difference in the nature of the fermions is that in Fermi superfluids and in superconductors
  the components of the Bogoliubov-Nambu spinor are related by complex conjugation. This reduces the number of degrees of freedom compared
  to Dirac case.
 }
\end{figure}

Fig. \ref{3He-B} shows the phase diagram of topological states of $^3$He-B in the plane $(\mu,1/m^*)$.
The line  $1/m^*=0$ corresponds to the Dirac vacuum of massive fermions, whose topological charge  is determined by the sign of mass parameter $M=-\mu$:
\begin{equation}
N_K= {\rm sign}(M)
\,.
\label{eq:DiracInvariants}
\end{equation}
Finally, the point $\mu=1/m^*=0$ corresponds to the massless Dirac particle, where the Dirac node consists of two Weyl nodes with opposite chirality,  $N_a=\pm N_d$.

The real superfluid $^3$He-B lives in the  corner of the phase diagram  $\mu>0$, $m^*>0$, $\mu\gg m^*c_B^2$, which also corresponds to the limit $\Delta_B\ll \mu$ of the weakly interacting gas of  $^3$He atoms,
where the superfluid state is described by Bardeen-Cooper-Schrieffer (BCS) theory. However, in the ultracold Fermi gases with triplet pairing
 the strong coupling limit is possible near the Feshbach resonance.\cite{GurarieRadzihovsky2007}
 When $\mu$ crosses zero the topological quantum phase transition occurs, at which the topological charge $N_K$ changes from  $N_K=2$ to  $N_K=0$. The latter regime with trivial topology also includes
 the Bose-Eistein condensate (BEC) of two-atomic molecules. In other words, the BCS-BEC crossover in this system is always accompanied by the topological quantum phase transition, at which the topological invariant changes.
 
There is an important difference between $^3$He-B and Dirac vacuum. The space of the Green's function of free  Dirac fermions is non-compact: $G$ has different asymptotes at $|{\bf p}|\rightarrow \infty $ for different directions of momentum ${\bf p}$.   As a result, the topological charge of the interacting Dirac fermions depends on the regularization at large momentum. $^3$He-B can serve as regularization of the Dirac vacuum, which can be made in the Lorentz invariant way.\cite{Volovik2010a} One can see from Fig. \ref{3He-B}, that  the topological charge of free Dirac vacuum has intermediate value between the charges of the  $^3$He-B vacua with compact  Green's function. On the marginal behavior of free Dirac fermions see Refs. \cite{Haldane1988,Schnyder2008,Volovik2003,Volovik2009b}.

   The vertical axis separates the states with the same asymptote of the Green's function at infinity. The abrupt change of the topological charge across the line, $\Delta N_K=2$, with fixed asymptote shows that one cannot cross the transition  line adiabatically. This means that all the intermediate states on the line of this  QPT  are necessarily gapless. For the intermediate state between the free Dirac vacua with opposite mass parameter 
 $M$ this is well known. But this is applicable to the general case with or without relativistic invariance:
 the gaplessness is protected 
by the difference of topological invariants on two sides of transition.
The gaplessness of the intermediate state leads also to the fermion zero modes at the interface between the bulk states with different topological invariants, see Sec. \ref{EdgeStates}.
For electronic materials this was discussed in Ref. \cite{VolkovPankratov1985}.

\section{ Combined topology. Evolution of Weyl points.}

According to Eq.(\ref{EandB}), the  effective (synthetic) electric and magnetic fields acting on the Weyl quasiparticles
emerge if the position ${\bf K}_a$ of the $a$-th  Weyl points depends on coordinate and time.
The topological protection of the Weyl points together with topology of the spatial distribution of the Weyl points in the coordinate space gives rise to the more complicated combined topology in the extended phase space $({\bf p},{\bf r})$.\cite{VolovikMineev1982} This combined topology connects the effect of chiral anomaly and the dynamics of skyrmions, which allowed us to observe experimentally the consequence of the chiral anomaly and to verify the Adler-Bell-Jackiw equation.\cite{BevanNature1997} We consider two examples of such dependence: the skyrmion in $^3$He-A and the core structure of the $^3$He-B vortex with ${\cal N}=1$, which exists at high pressure.

\begin{figure}
 \includegraphics[width=0.5\textwidth]{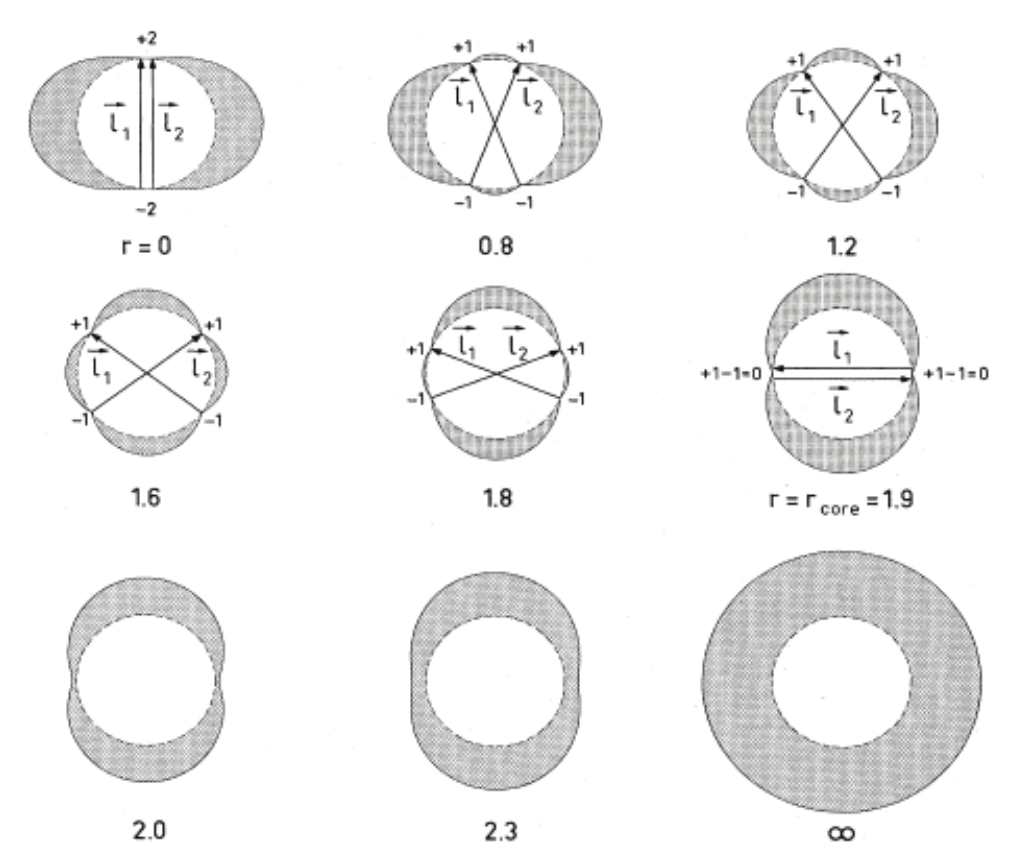}
 \caption{Combined topology of the Weyl points in the  core of the axisymmetric $^3$He-B vortex
according to Ref. \cite{VolovikMineev1982} (from review paper \cite{SalomaaVolovik1987}). Here $r$ is the distance from the vortex axis. Weyl points are topologically stable nodes in the quasiparticle particle spectrum, 
which have integer topological charge  $N$ in momentum space. The  Weyl points are situated at momenta 
${\bf K}_a$  in momentum space, which depend on the position ${\bf r}$ in the real space.
In superfluid $^3$He, the Weyl points live at $p=p_F$, i.e. on the former Fermi surafce, ${\bf K}_a=\pm p_F\hat{\bf l}_a$, where $\hat{\bf l}_a$ are unit vectors. That is why originally the Weyl point was called "boojum on Fermi surface".   On the vortex axis, at $r=0$, one has two pairs of Weyl points with $\hat{\bf l}_1=\hat{\bf l}_2=\hat{\bf z}$. Each pair forms the Weyl point with double topological charges, $N=+2$ on the north pole and  $N=-2$ on the south pole.
This corresponds to the chiral $^3$He-A on the axis without any vorticity.
For $r>0$, the multiple nodes split into pairs of Weyl points, each carrying unit topological charges $N=+1$ or $N=-1$. For increasing $r$, the Weyl points move continuously towards the
equatorial plane, where they annihilate each other ($+1-1=0$). For larger $r$ the fully gapped state is formed, which becomes the isotropic $^3$He-B  far from the vortex.
The coordinate dependence of the Weyl point gives rise to vorticity concentrated in the vortex core, as a result the vortex in the B-phase acquires the winding number. In other words, according to Ref. \cite{VolovikMineev1982} the vortex -- the topological defects  in ${\bf r}$-space --  flows out into ${\bf p}$-space due to evolution of the Weyl points. The topology of the evolution is governed by Eq.(\ref{Combined}), which connects three topological invariants: real-space winding number of the vortex ${\cal N}$, momentum-space invariant of the Weyl point $N$ and the invariant $\nu$, which describes the evolution of the Weyl point in real space.
 }
 \label{Evolution}
\end{figure}

\subsection{From A to B. Topology of Weyl nodes in $^3$He-B vortex.}
\label{AtoB}

 In the core of the  $^3$He-B axisymmetric vortex with the spontaneously broken parity\cite{MagneticVortices1983} one has the  $^3$He-A order parameter on the vortex axis, at $r=0$, which continuously transforms to the  $^3$He-B order parameter far from the core. 
Fig. \ref{Evolution} demonstrates the evolution of the Weyl points on the way from $^3$He-A to the $^3$He-B.\cite{VolovikMineev1982,SalomaaVolovik1987}  
 At $r>0$ the spin degeneracy of the Weyl points is lifted, and the nodes with $N=\pm 2$ split into the  elementary Weyl nodes with $N=\pm 1$.
  At $r=r_{\rm core}$ the Weyl points with opposite $N$  merge to form the Dirac points with trivial topological charge, $N=0$.  At $r>r_{\rm core}$, the Dirac points disappear, because they are not protected by topology,  and the fully gapped state emerges.  Far from the vortex core one obtains the  $^3$He-B order parameter with the $2\pi$ phase winding around the vortex line.

In this evolution of Weyl points the  chirality of $^3$He-A, which is the property of topology in momentum space,  continuously transforms to the integer valued circulation of superfluid velocity around the vortex, which is described by the real space topology.
The topological connection of the real-space and momentum-space properties is encoded in equation (4.7) of Ref.\cite{VolovikMineev1982}:
\begin{equation}
{\cal N} =\frac{1}{2}\sum_a N_a\nu_a \,.
\label{Combined}
\end{equation}
Here ${\cal N}$ is the real-space topological invariant -- the winding number of the vortex;
$N_a$ is the momentum-space topological invariant describing the $a$-th Weyl point.
Finally the index $\nu_a$ connects the two spaces: it shows how many times the Weyl point ${\bf K}_a$  covers sphere, when the coordinates ${\bf r}=(x,y)$ run over the cross-section of the vortex core:
\begin{equation}
\nu_a=\frac{1}{4\pi} \int dxdy \frac{1}{|{\bf K}^a|^3}{\bf K}^a\cdot(\partial_x {\bf K}^a \times \partial_y {\bf K}^a)\,.
\label{nu}
\end{equation}

For the discussed $^3$He-B vortex the Weyl nodes with $N_a=\pm 1$ cover the half a sphere,  $\nu_a=\pm 1/2$, which gives ${\cal N} =\frac{1}{2}(1/2 + 1/2+ 1/2+1/2)=1$.
 
\subsection{Topology of evolution of the Weyl nodes in the $^3$He-A skyrmion.}

Eq.(\ref{Combined}) is also applicable to the continuous textures  in $^3$He-A.  For example, in the skyrmion in Fig. \ref{Skyrmion2merons} the Weyl nodes with $N_a=\pm 2$ cover the whole sphere once,  $\nu_a=\pm 1$. This gives ${\cal N} =\frac{1}{2}(2\times 1 + (-2)\times(-1))=2$, which means that  the skyrmion represents the continuous doubly quantized vortex. The steps in the NMR spectrum corresponding to the ${\cal N}=2$ vortices were observed  in the NMR experiments on rotating $^3$He-A.\cite{Blaauwgeers2000}   The meron -- the Mermin-Ho vortex with $N_a=\pm 2$ and 
 $\nu_a=\pm 1/2$ -- represents the singly quantized vortex, ${\cal N} =\frac{1}{2}(2\times \frac{1}{2} + (-2)\times (-\frac{1}{2}))=1$. The elementary cell of the skyrmion lattice in the low magnetic field contains 4 merons, see Fig. \ref{FourMerons}, and thus has ${\cal N}=4$. Merons are also the building blocks of the vortex sheet observed in $^3$He-A.\cite{Parts1994a,Parts1994b}

\subsection{Topology of the phase of the gap function in $({\bf p},{\bf r})$-space.}

The close connection between topologies in real and momentum space can be also seen when equation (\ref{Determinant}) for the gap function is extended to the inhomogeneous case
\begin{equation}
{\rm det}\,\hat\Delta({\bf p},{\bf r})=|{\rm det}\,\hat\Delta({\bf p},{\bf r})| e^{i\Phi({\bf p},{\bf r})} \,.
\label{DeterminantExtended}
\end{equation}
Then the phase $\Phi({\bf p},{\bf r})$ of the determinant of the  gap function may include the phase winding in real space (the state with quantized vortex) and the winding in momentum space, which gives rise to the line of zeroes in the determinant of the gap function and thus to the Weyl points in spectrum.  In general, the winding  number of the phase protects the 4-dimensional vortex singularity of the phase $\Phi({\bf p},{\bf r})$  in the $(3+3)$-dimensional $({\bf p},{\bf r})$-space. By changing the orientation of this 4-D manifold in the $(3+3)$-space, one can transform the  $^3$He-B state with the ${\cal N}=1$ vortex,
to the homogeneous $^3$He-A state  with Weyl points. In the vortex state of $^3$He-B, outside of the vortex core the phase depends only on the coordinates:
$\Phi({\bf p},{\bf r})=\Phi(\phi)=2\phi$. The corresponding  4D manifold is the $(3+1)$-subspace (1D vortex line times the 3D momentum space). In the vortex-free  $^3$He-A state the phase depends only on the momentum, 
$\Phi({\bf p},{\bf r})=\Phi({\bf p})$, and it has the winding number $N$ in momentum space.
In this case the corresponding  4D manifold is the $(1+3)$-subspace (the 1D line of determinant nodes in momentum space times the 3D coordinate space).

For the inhomogeneous $^3$He-A with $\hat{\bf l}({\bf r})$,  the nonzero winding of the phase $\Phi({\bf p},{\bf r})$ gives in particular the following 4D generalization of the singular vorticity:\cite{1981,VolovikMineev1982}
\begin{eqnarray}
\left(
 \frac{\partial}{\partial {\bf p}} \cdot  \frac{\partial}{\partial {\bf r}} -  \frac{\partial}{\partial {\bf r}} \cdot  \frac{\partial}{\partial {\bf p}}
 \right)  \Phi({\bf p},{\bf r})=
\nonumber
\\
 -2\pi\, (\hat{\bf l}\cdot {\bf p})\, (\hat{\bf l} \cdot \nabla \times \hat{\bf l}) 
 \, \delta\left( {\bf p}_\perp({\bf r})\right)\,,
\label{Singularity1}
 \\
 {\bf p}_\perp({\bf r})={\bf p}-\hat{\bf l}({\bf r}) (\hat{\bf l}({\bf r}) \cdot {\bf p})\,.
\label{Singularity2}
\end{eqnarray}

\section{Chiral anomaly}

\subsection{Hydrodynamic anomalies in chiral superfluids}

The singularity in Eq.(\ref{Singularity1}) leads to the anomalies in the equations for the mass current (linear momentum density) and angular momentum density of the chiral liquid, since these quantities can be expressed via the gradients of the generalized phase $\Phi({\bf p},{\bf r})$.\cite{1981,VolovikMineev1982} Later it became clear, that  these anomalies are the manifestation of the chiral anomaly in $^3$He-A related to the Weyl points, which led to the modifcation of the hydrodynamic equations derived by Khalatnikov and Lebedev \cite{KhalatnikovLebedev1977}. The effect of the chiral anomaly has been observed in experiments with dynamics of the 
vortex-skyrmions,\cite{BevanNature1997} which revealed the existence of the anomalous spectral-flow force acting on skyrmions. 

Originally the anomalies in the dynamics of $^3$He-A have been obtained from the calculations of the response functions and from the hydrodynamic equations, which take into account the singularity of the phase $\Phi$ 
in Eq.(\ref{Singularity1}).\cite{1981,VolovikMineev1982}  In these calculations the main contribution to the anomalous behavior comes from the momenta ${\bf p}$ far from the Weyl points, where the spectrum is highly nonrelativistic.
The results of calculations coincide with the later results obtained using the relativistic spectrum 
of chiral fermions emerging in the vicinity of  the Weyl points.  This is because the spectral flow through the Weyl nodes, which is in the origin of anomalies, does not depend on energy and is the same far from and close to the nodes.

The original approach uses the semiclassical approximation, which takes into accouns that the phase 
$\Phi({\bf p},{\bf r})$ of the determinant can be considered as the action in the quasiparticle dynamics,
\begin{equation}
S({\bf p},{\bf r}) =\frac{\hbar}{4} \Phi({\bf p},{\bf r})\,.
\label{ActionS}
\end{equation}
The mass current ${\bf j}=\sum_{\bf p}{\bf p}f({\bf p},{\bf r})$, where $f$ is the distribution function of bare particles (atoms of $^3$He), see details in \cite{1981,VolovikMineev1982}. In  the inhomogeneous state the momentum and coordinate are shifted by $\nabla S$ and $- \frac{\partial S}{\partial {\bf p}}$ respectrively, and one has at $T=0$   (we take $\hbar=1$):
\begin{eqnarray}
{\bf j}= \sum_{\bf p}{\bf p}f\left({\bf p}- \nabla S,{\bf r}+ \frac{\partial S}{\partial {\bf p}}\right)
=
\nonumber
\\
\frac{1}{4}\sum_{\bf p}{\bf p} \left(\nabla f \cdot\frac{\partial \Phi}{\partial {\bf p}} - \nabla \Phi \cdot \frac{\partial f}{\partial {\bf p}}\right)=
\label{SemiclassicalCurrent2}
\\
=\frac{1}{4}\sum_{\bf p} f\nabla\Phi + \frac{1}{4}\nabla_i \left(\sum_{\bf p}  {\bf p} f \frac{\partial \Phi}{\partial p_i}\right) - 
\nonumber
\\
-\frac{1}{4}\sum_{\bf p} {\bf p} f\left(
 \frac{\partial}{\partial {\bf p}} \cdot  \frac{\partial}{\partial {\bf r}} -  \frac{\partial}{\partial {\bf r}} \cdot  \frac{\partial}{\partial {\bf p}}
 \right)  \Phi({\bf p},{\bf r})=
\label{SemiclassicalCurrent2}
\\
= \rho {\bf v}_{\rm s} + \frac{1}{2}\nabla \times {\bf L}
-\frac{1}{2}C_0 \hat{\bf l} (\hat{\bf l}\cdot \nabla \times \hat{\bf l}) 
\,.
   \label{Current}
\end{eqnarray}
The first  term in Eq.(\ref{Current}) is the superfluid mass current with velocity ${\bf v}_{\rm s}$ and mass density $\rho=mn$, where $n$ is particle density.
The second term is the mass current produced by the inhomogeneity of the orbital angular momentum density 
${\bf L}=(\hbar n/2) \hat{\bf l}$ in the chiral liquid. This is what one would naturally expect for the angular momentum of the chiral liquid with  particle density $n$ and the angular momentum $\hbar\hat{\bf l}$ for each pair of atoms.
The last term in Eq.(\ref{Current}) is anomalous, it is nonzero due to the 4D vortex singularity in the gap function determinant in Eq.(\ref{Singularity1}).
 The parameter $C_0$, which characterizes this hydrodynamic anomaly, equals:
\begin{equation}
 C_0=\frac{p_F^3}{3\pi^2\hbar^2} \,.
\label{C0}
\end{equation}
The same parameter enters the other hydrodynamic anomalies in the chiral superfluid: in the conservation laws for 
the linear momentum and the angular momentum  in chiral superfluids. At $T=0$ one has:\cite{1981} 
\begin{eqnarray}
\partial_t j_i -\nabla_k\pi_{ik} = -\frac{3}{2}C_0 \hat l_i (\partial_t\hat{\bf l}\cdot \nabla \times \hat{\bf l}) \,,
\label{nonconservation}
\\
\frac{\partial {\bf L}}{\partial t} + \frac{\delta E}{\delta  {\mbox{\boldmath$\theta$}}}  
=\frac{1}{2}C_0 \frac{\partial\hat{\bf l}}{\partial t}
 \,.
\label{L}
\end{eqnarray}
The nonzero value of the rhs  of Eq.(\ref{nonconservation}) manifests the non-conservation of the vacuum current, which means that the linear momentum is carried away by the fermionic quasiparticles created from the superfluid vacuum. The Eq.(\ref{L}) shows that the angular momentum of the superfluid vacuum is also not conserved. This equation can be also represented in terms of the nonlocal  variation of the angular momentum:
$\partial_t \delta {\bf L}= - \delta E/\delta  {\mbox{\boldmath$\theta$}}$ with $\delta {\bf L}=\frac{1}{2}  (n -C_0) \delta\hat{\bf l}  -\frac{1}{2}\hat{\bf l}  \delta n$. In other words, there is the dynamical reduction of the angular momentum from its static value $\frac{\hbar}{2}  n \hat{\bf l}$   to the dynamic value $\frac{\hbar}{2}  (n - C_0)\hat{\bf l}$. 
Note that in $^3$He-A the Cooper pairing is in the weak coupling regime, which means that the gap amplitude $\Delta$ is much smaller that the Fermi energy $\mu$. As a result the particle density $n$ in the superfluid state is very close to the parameter $C_0$, which is equal to the particle density in the normal state, $C_0=n(\Delta=0)$. One has $(n-C_0)/C_0 = (n(\Delta)-n(\Delta=0))/n(\Delta=0)\sim 10^{-5}$, so that the reduction of the dynamical angular momentum is crucial. 

\subsection{Hydrodynamic anomalies from chiral anomaly}

To connect the hydrodynamic anomalies in Eqs.(\ref{nonconservation})-(\ref{L}) and the chiral anomaly in relativistic theories, let us take into account that the parameter $p_F$ which enters $C_0$ marks the position of the Weyl points in $^3$He-A:
${\bf K}_a=\pm p_F\hat{\bf l}$. When $\mu\rightarrow 0$ and correspondingly ${\bf K}_a\rightarrow 0$, the Weyl points merge and annihilate. At the same time  $C_0\rightarrow 0$, and all the hydrodynamic anomalies disappear. They do not exist in the strong coupling regime at $\mu<0$, where the chiral superfluid has no Weyl points and $C_0=0$.  All the dynamic anomalies experienced by $^3$He-A come from the Weyl points: the exitence of the Weyl nodes in the spectrum  allows the spectral flow of the fermionic levels through the nodes, which carry the linear and angular momentum  from the vacuum to the quasiparticle world. The state of the chiral superfluid with anomalies at $\mu>0$ and the anomaly-free state of the chiral superfluid at $\mu<0$ are separated by the topological quantum phase transition at $\mu=0$.

Close to the Weyl point the spectral flow can be considered in terms of the relativistic fermions.
The chiral fermions experience the effect of chiral anomaly in the presence of the synthetic 
electric and magnetic fields in Eq.(\ref{EandB}).  The left-handed or right-handed fermions are created according
to the Adler-Bell-Jackiw equation for chiral anomaly:
\begin{equation}
\dot n_R =- \dot n_L=\frac{1}{4\pi^2}  q^2{\bf B}_{\rm eff} ({\bf r},t)\cdot {\bf E}_{\rm eff} ({\bf r},t)
\,.
\label{ABJ}
\end{equation}
The fermionic quasiparticles created from the superfluid vacuum carry the fermionic charge from the vacuum to the "matter" -- the normal component of the liquid, which at low temperatures consists of thermal Weyl fermions. For us the important fermionic charge is the quasiparticle momentum: each fermion created from the vacuum carry with it the momentum ${\bf K}^{(a)}= \pm p_F\hat{\bf l}$. According to the Adler-Bell-Jackiw equations for chiral anomaly, this gives the following momentum creation from the vacuum per unit time per unit volume:
\begin{equation}
\partial_t P_i -\nabla_k\pi_{ik} =\frac{1}{4\pi^2}  {\bf B}_{\rm eff} ({\bf r},t)\cdot {\bf E}_{\rm eff} ({\bf r},t)\sum_a   K_i^{(a)} N_a q_a^2 
\,.
\label{MomentumProductionGeneral}
\end{equation}
Here as before $N_a$ is the topological charge of the $a$-th Weyl point, which sign determines the chirality of the Weyl quasiparticles near the Weyl node; and $q_a=\pm 1$ is the effective electric charge. The rhs of Eq.(\ref{MomentumProductionGeneral}) is non-zero, because the quasiparticles with opposite chirality $N_a$ carry opposite momentum ${\bf K}^{(a)}$.
Since in the supefluids the momentum density equals the mass current density, ${\bf P}={\bf j}$, the Eq.(\ref{MomentumProductionGeneral}) reproduces the equation (\ref{nonconservation}). This demonstrates that nonconservation of the linear momentum of the superfluid vacuum is the consequence of the chiral anomaly.

The angular momentum anomaly is also related to the spectral flow through the nodes in  bulk or on the surface.\cite{Volovik1995,Tada2015,Volovik2015,Ojanen2016,Shitade2016}

\begin{figure}
 \includegraphics[width=0.5\textwidth]{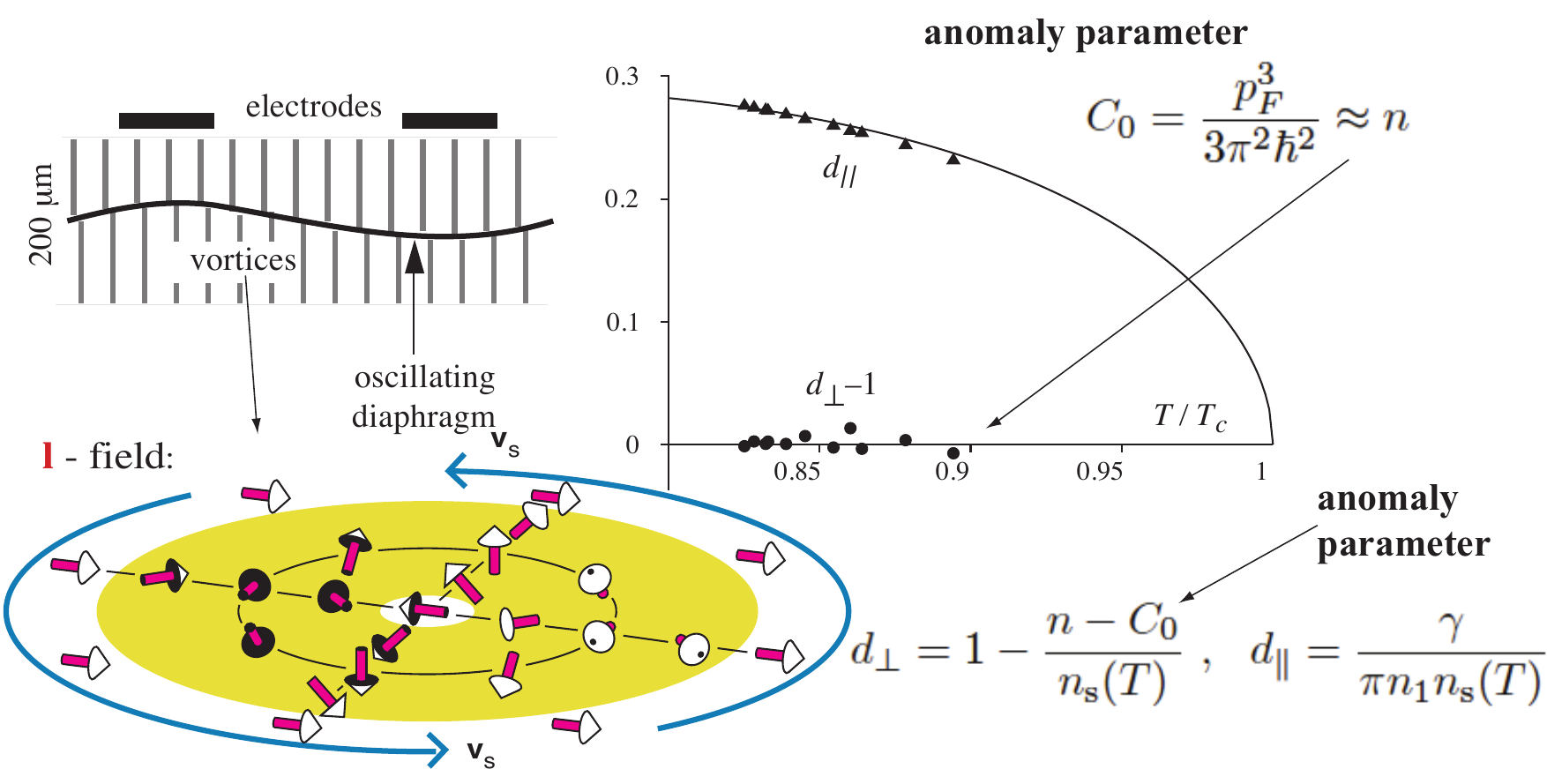}
 \caption{
Experimental verification of the Adler-Bell-Jackiw anomaly equation in $^3$He-A
 (from Ref.\cite{BevanNature1997}).
{\it Left}: A uniform array of vortices is produced
by rotating the whole cryostat, and oscillatory superflow perpendicular
to the rotation axis is produced by a vibrating diaphragm, while
the normal fluid (thermal excitations) is clamped by viscosity, ${\bf v}_{\rm
n}=0$. The velocity
${\bf v}_{\rm L}$ of the vortex array is determined by the overall
balance of forces acting on the vortices in Eq.(\ref{ForceBalance}).  The
vortices
produce the dissipation $d_\parallel$ and also the
coupling between two orthogonal modes of membrane oscillations, which is proportional to
$d_\perp$.  {\it Right}: The parameters $d_\parallel$ and $d_\perp$ measured for  the continuous vortices-skyrmions  in $^3$He-A. As distinct from the $^3$He-B vortices, for skyrmions the measured parameter  $d_\perp$ is close to unity. According to Eq.(\ref{d_perpAphase}), the experiment demonstrates that the anomaly parameter $C_0$ is close to the particle density $n$ and thus it verifies the Adler-Bell-Jackiw anomaly equation (\ref{MomentumProductionGeneral}).
 }
 \label{Bevan}
\end{figure}

According to the Newton law, the creation of the linear momentum from the vacuum per unit time due to the spectral flow through the Weyl nodes under the effective electric and magnetic fields produced by  the time dependent texture of the vector $\hat{\bf l}({\bf r},t)$, is equivalent to an extra force acting on the texture. In experiment, the relevant  time dependent texture 
is the vortex-skyrmion moving with velocity ${\bf v}_{\rm L}$, where 
$\hat{\bf l}({\bf r},t)=\hat{\bf l}({\bf r} -{\bf v}_{\rm L} t)$. This together with the effective magnetic field 
${\bf B}_{\rm eff}=p_F \nabla\times \hat{\bf l}$ gives also the effective electric field 
${\bf E}_{\rm eff}=-p_F \partial_t\hat{\bf l}=p_F( {\bf v}_{\rm L} \cdot \nabla) \hat{\bf l}$.

 The anomalous spectral-flow force acting on this topological object  is obtained after integration of the rhs of Eq.(\ref{MomentumProductionGeneral}) over the cross-section of the skyrmion:
\begin{eqnarray}
{\bf F}_{\rm ~spectral~flow}= 
\nonumber
\\
\frac{p_F^3}{2\pi^2} \int d^2x  \, \hat{\bf l} \left( (\nabla\times \hat{\bf l} )\cdot  ( {\bf v}_{\rm L} \cdot \nabla) \hat{\bf l}\right) = - \pi {\cal N} C_0 {\hat {\bf z}}  \times {\bf v}_{\rm L} \,.
\label{KopninForce}
\end{eqnarray} 
Here ${\cal N}$ is the vortex winding number of the texture, which via Eq.(\ref{Combined}) 
is expressed in terms of the Weyl point charges $N_a$ and  the topological $\pi_2$ charges of their spatial distributions in the texture. This demonstrates that the spectral flow force acting on
the vortex-skyrmion, which has ${\cal N}=2$, is the result of the combined effect of real-space and momentum-space topologies.

\subsection{Experimental observation of chiral anomaly in chiral superfluid $^3$He-A}

The force acting on the vortex-skyrmions has been measured experimentally.\cite{BevanNature1997}
There are several forces acting on vortices, including the Magnus force, Iodanski force  and the anomalous spectral-flow force, which is called the Kopnin force. For the steady state motion of vortices the sum of all forces
acting on the vortex must be zero. This gives the following equation connecting velocity of the
superfluid vacuum ${\bf v}_{\rm s}$, the velocity of the vortex line ${\bf v}_{\rm L}$ and the velocity ${\bf v}_{\rm n}$ of "matter" -- the velocity of the normal component of the liquid:
\begin{equation}
\hat{\bf z}\times ({\bf v}_{\rm L}-{\bf v}_{\rm s})+
d_\perp \hat{\bf z}\times({\bf
v}_{\rm n}-{\bf v}_{\rm L})+d_\parallel({\bf v}_{\rm n}-{\bf v}_{\rm L})
=0\,.
\label{ForceBalance}
\end{equation}
For the continuous vortex-skyrmion in $^3$He-A with the spectral flow force in Eq.(\ref{KopninForce}) the reactive parameter $d_\perp$ is expressed in terms of the anomaly parameter $C_0$:
 \begin{equation}
d_\perp-1= \frac{C_0-n}{n_{\rm s}(T)}\,.
\label{d_perpAphase}
\end{equation}
Here $n_{\rm s}(T)=n- n_{\rm n}(T)$ is the density of the superfluid component.

Since in $^3$He-A the anomaly parameter $C_0$ is very close to the particle density $n$, the chiral anomaly in $^3$He-A should lead to equation $d_\perp-1=0$ for practically all temperatures. This is what has been observed in Manchester experiment on skyrmions in $^3$He-A, 
see Fig. \ref{Bevan} ({\it right}) which experimentally confirms
the generalized Adler-Bell-Jackiw equation (\ref{MomentumProductionGeneral}). 

In conclusion, the chiral anomaly related to the Weyl fermionic quasiparticles, whose gapless spectrum is protected by the topological invariant in ${\bf p}$-space, has been observed in the experiments with skyrmions --  objects, which are protected by the topological invariant in the ${\bf r}$-space.
The effect of chiral anomaly observed in $^3$He-A incorporates several topological charges described by the combined topology in the extended $({\bf p},{\bf r})$-space, which is beyond the conventional anomalies in the relativistic systems. 

\subsection{Chiral magnetic and chiral vortical effects in $^3$He-A}

\begin{figure}
 \includegraphics[width=0.5\textwidth]{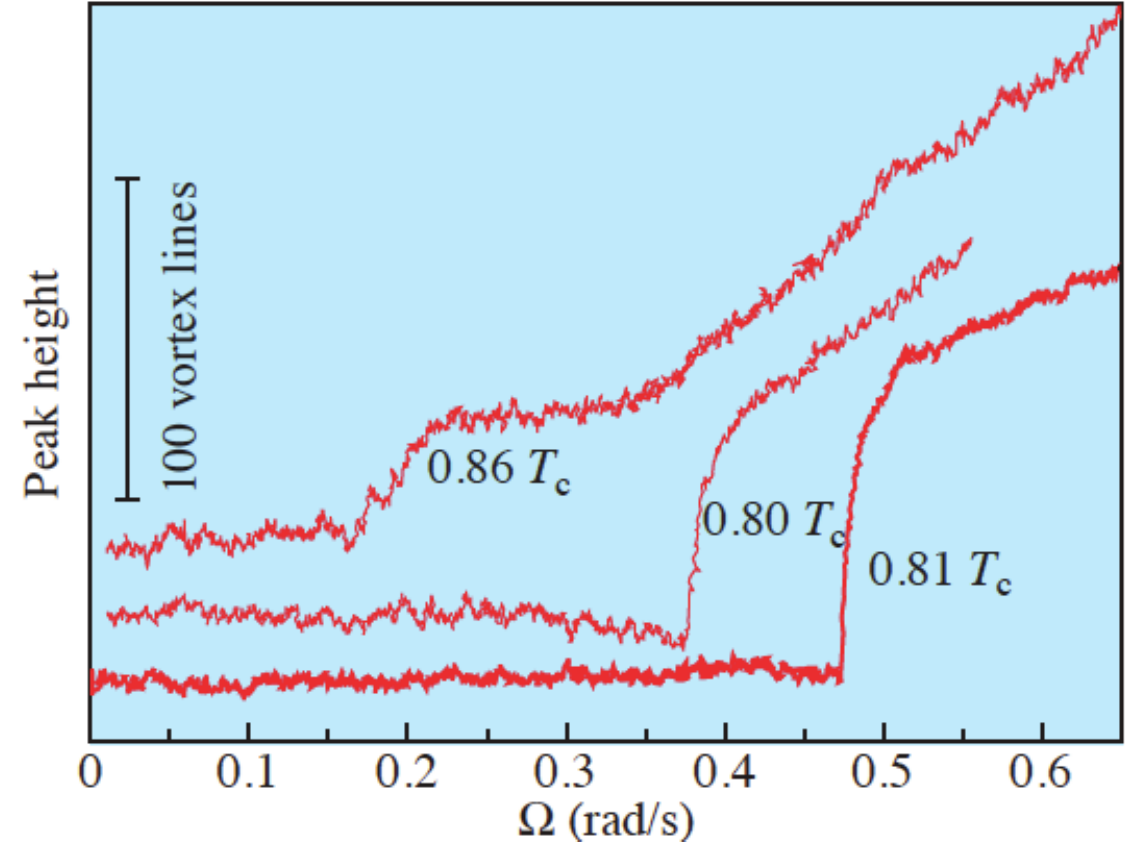}
 \caption{Demonstration of the chiral magnetic effect in $^3$He-A.\cite{ExperimentalMagnetogenesis1997} 
In NMR experiments the height of the satellite peak is measured, which comes from  
the vortex skyrmions, see Fig. \ref{Skyrmion2merons}c. Initially no vortices are present in the vessel.
When the velocity of the superflow ${\bf v}_{\rm s}$, which corresponds
to the chiral chemical potential in Eq.(\ref{CME}),
exceeds a critical value determined by spin-orbit interaction, the helical instability takes place, and
the container becomes filled with the skyrmions. Skyrmions carry the analog of a
hypermagnetic field, as a result the process of their creation, which is governed by chiral anomaly, is analogous to the process of formation of magnetic field
in early Universe.\cite{Joyce1997,Shaposhnikov1997} 
}
 \label{Magnetogenesis}
\end{figure}

Another combination of the fermionic charges of the Weyl fermions gives rise to the chiral magnetic effect (CME) -- the topological mass current along the magnetic field. The effect can be written in terms of the following contribution to the free energy:
\begin{equation}
F_{\rm CME}=\frac{1}{8\pi^2} \int d^3x\,  {\bf A}_{\rm eff} ({\bf r},t)\cdot {\bf B}_{\rm eff} ({\bf r},t)
\sum_a   \mu^{(a)} N_a q_a^2 
\,,
\label{CME}
\end{equation}
where $\mu^{(a)}$ are chemical potentials of the right and left Weyl quasiparticles.
The variation of the energy over the effective vector potential  ${\bf A}_{\rm eff}=p_F \hat{\bf l}$ gives the effective current along the effective magnetic field. The CME is nonzero if there is a disbalance of left and right quasiparticles. In $^3$He-A this disbalance is achieved by application of supercurrent,
which produces the chemical potentials for chiral quasiparticles with opposite sign:
$\mu^{(a)} = \pm p_F \hat{\bf l}\cdot {\bf v}_{\rm s}$, see Refs. \cite{KrusiusVachaspatiVolovik1998,Volovik2003}
for details. The supercurrent is created in the rotating cryostat.
For us the most important property of the CME term is that it is linear in the gradient  of $\hat{\bf l}$. Its sign thus can be negative, which leads to the observed helical instability of the superflow towards formation of the inhomogeneous $\hat{\bf l}$-field in the form of skyrmions, Fig. \ref{Magnetogenesis}.\cite{ExperimentalMagnetogenesis1997} 
Since the texture $\nabla \times \hat{\bf l}$ inside the skyrmion plays the role of magnetic field, the process of formation of skyrmions
in the superflow is analogous to the formation of the (hyper)magnetic fields in the early Universe.\cite{Joyce1997,Shaposhnikov1997} 
Now the CME is studied in relativistic heavy ion collisions where strong
magnetic fields are created by the colliding ions \cite{Kharzeev2016}.

\begin{figure}
 \includegraphics[width=0.5\textwidth]{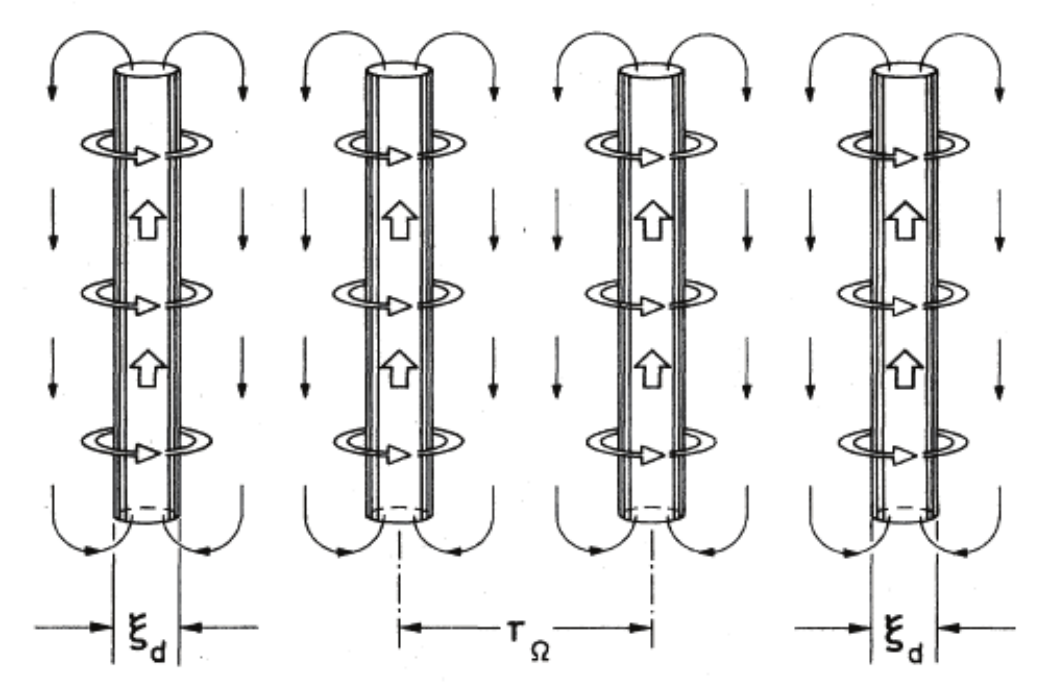}
 \caption{Schematic illustration of the chiral vortical effect in rotating $^3$He-A from  Ref.\cite{SalomaaVolovik1987}.
The mass currents along the rotation axis, which are concentrated in the soft cores of the vortex skyrmions, are fully compensated by the bulk current in the equilibrium state.
This is the realization of the Bloch theorem which forbids the total current in the condensed matter system in equilibrium.
}
 \label{VorticalEffect}
\end{figure}

The related effect is the chiral vortical effect (CVE), see e.g. review \cite{Zubkov2018}. It can be described by the following term in the free energy \cite{Volovik2003}:
\begin{equation}
F_{\rm CVE}=\frac{1}{8\pi^2} \int d^3x\,  {\bf A}_{\rm eff} ({\bf r},t)\cdot {\bf B}_{g} ({\bf r},t)
\sum_a   (\mu^{(a)})^2  N_a q_a
\,.
\label{CVE}
\end{equation}
Here ${\bf B}_{g}$ is the effective gravimagnetic field, which in $^3$He-A  is produced for example by rotation, ${\bf B}_{g}=2 {\mbox{\boldmath$\Omega$}}/c^2$, where ${\mbox{\boldmath$\Omega$}}$ is the angular velocity of rotation. The variation of the energy over the effective vector potential  ${\bf A}_{\rm eff}=p_F \hat{\bf l}$ gives the effective current along the rotation axis.
The real mass current along the rotation axis has been discussed in Ref.\cite{SalomaaVolovik1987}, see Fig. \ref{VorticalEffect}.
 
In general, the total current along the magnetic field or along the rotation axis in the ground state of the condensed matter system is prohibited by 
the Bloch theorem \cite{Yamamoto2015,Watanabe2019}.
In our case of CME, the field ${\bf A} =  p_F\hat{\bf l}$ is effective, and the corresponding current 
${\bf J}=\delta  F/\delta {\bf A}$ is also effective.  It does not coincide with the real mass current 
${\bf j}=\delta  F/\delta {\bf v}_{\rm s}$. The imbalance between the chiral chemical potentials of left-handed and right-handed Weyl fermions is also effective: it is provided by the superflow due to the Doppler shift.  For the effective fields and currents the no-go theorem is not applicable.

In the case of the chiral vortical effect with the real mass current along the rotation in Fig. \ref{VorticalEffect} the Bloch theorem is obeyed.  Such configuration corresponds to the ground state in the given topological class of rotating states
in $^3$He-A. In equilibrium, the total mass current along the rotation axis is absent: the currents concentrated in the soft cores of the vortex-skyrmions are compensated by the opposite superfluid current in bulk.

\section{Fermion zero modes}

The nontrivial topology of $^3$He superfluids leads to the topologically protected massless (gapless) Majorana fermions living on the surface of the superfluid or/and inside the vortex core.
Let us start with the fermion zero modes on the surface of $^3$He-B.

\subsection{Majorana edge states in $^3$He-B}
\label{EdgeStates}

\begin{figure}
 \includegraphics[width=0.4\textwidth]{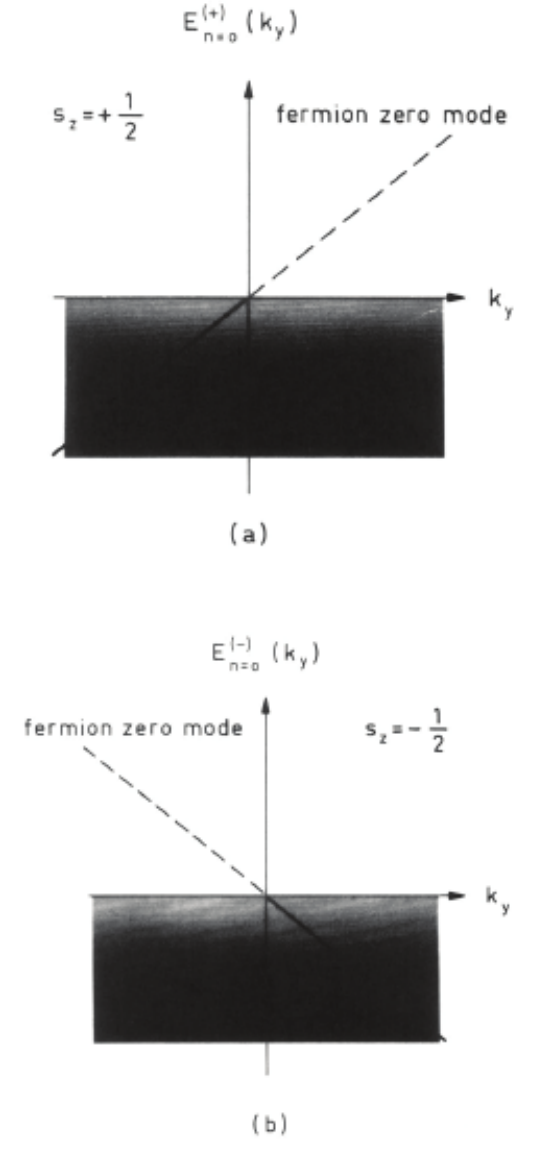}
 \caption{Schematic illustration of the spectrum of Majorana fermions at the Kibble wall in $^3$He-B -- the nontopological domain wall
which separates the regions with the opposite topological charges $N_K=+2$ and $N_K=-2$ (from 
Ref.\cite{SalomaaVolovik1988}).  The spectrum satisfies Eq.(\ref{eq:ModesH}).
 }
 \label{MajoranaB}
\end{figure}

 Fig.\ref{MajoranaB} reproduces the Fig. 12 from Ref.\cite{SalomaaVolovik1988} for the fermion zero modes living 
at the interface between two bulk states of $^3$He-B with opposite topological charges $N_K$ in Eqs.(\ref{dinvariant})-(\ref{3DTopInvariant_tau}).
There are two branches $E(p_y,p_z)$ of  fermion zero modes with different directions of spin, which form the Majorana cones. They are described by the effective 2+1 theory with the following Hamiltonian:
 \begin{equation}
H = c \hat{\bf x} \cdot({\mbox{\boldmath$\sigma$}} \times {\bf p})\,.
\label{eq:ModesH}
\end{equation} 
Here $\hat{\bf x}$ is along the normal to the interface, and $c$  is the "speed of light" of the emergent relativistic spectrum of these surface fermions. The same Hamiltonian describes the surface states on the boundary of $^3$He-B,\cite{ChungZhang2009,Volovik2009b} which we consider here.

The parameter $c$ depends on the structure of the  anisotropic order parameter $A_{\alpha i}(x)$ in the interface, or near the wall of the container. Let us consider the order parameter near the wall, with the bulk $^3$He-B at $x>0$:
\begin{equation}
A_{\alpha i}= \left(
\begin{matrix}
 \Delta_\perp(x)&0&0\cr
0&\Delta_\parallel(x)&0\cr
0&0&\Delta_\parallel(x)\cr
\end{matrix} 
\right) \,.
\label{OrderParameter}
\end{equation}
 $\Delta_\perp(x=\infty)=\Delta_\parallel(x=\infty)=\Delta_B$, where  $\Delta_B$ is the gap in bulk $^3$He-B.
The corresponding 3D Hamiltonian is:
\begin{eqnarray}
H=\frac{p^2-p_F^2}{2m^*} \tau_3+
\nonumber
\\
   \tau_1 \left(\Delta_\perp(x)\sigma_x \frac{p_x}{p_F} +\Delta_\parallel(x) \sigma_y  \frac{p_y}{p_F}+ \Delta_\parallel(x)\sigma_z  \frac{p_z}{p_F}  \right)\,,
\label{eq:Hamiltonian}
\end{eqnarray}
where $\tau_\alpha$ and $\sigma_i$ are as before the Pauli matrices of Bogoliubov-Nambu spin  and nuclear spin correspondingly; and $p_x$ now is the operator. Let us find the bound surface states of this Hamiltonian.

We can use the method of trajectories. In $^3$He superfluids the Fermi momentum $p_F \gg 1/\xi$, where $\xi$ is   the coherence length, which determines the characteristic thickness of the surface layer, where the order parameter evolves. That is why with a good accuracy the classical trajectories are the straight lines, and we can consider the Hamiltonian along the trajectory.
We assume here  the specular reflection of quasiparticles at the boundary, then the momentum component $p_x$ which is normal to the wall of the container, changes sign after reflection.  This means that after reflection from the wall the quasiparticle moving along the trajectory feels the change from $\Delta_\perp(x)$ to $- \Delta_\perp(x)$. Then the problem  transforms to that discussed in  Ref.\cite{SalomaaVolovik1988} of finding the spectrum of the fermion bound states at the interace separating
two bulk $^3$He-B states  with the opposite values of the topological charge $N_K=\pm 2$ in Eq.(\ref{MasslessTopInvariant3D}):
\begin{equation}
A_{\alpha i}(+\infty)= \left(
\begin{matrix}
 \Delta_B&0&0\cr
0&\Delta_B&0\cr
0&0&\Delta_B\cr
\end{matrix} 
\right)~~,~~ N_K=+2 \,,
\label{OrderParameterInterface+}
\end{equation}
and 
\begin{equation}
A_{\alpha i}(-\infty)=
 \left(
\begin{matrix}
 -\Delta_B&0&0\cr
0&\Delta_B&0\cr
0&0&\Delta_B\cr
\end{matrix} 
\right) ~~,~~ N_K=-2 \,.
\label{OrderParameterInterface-}
\end{equation}

Introducing $p_x=p_F -i\partial_x$ and neglecting the terms  quadratic in  $p_y$, $p_z$ and $\partial_x$ one obtains the Hamiltoinian
 \begin{eqnarray}
H=H_0+H_1~,
\label{eq:HamiltonianExpansion}
\\
H_0=-i v_F \tau_3\partial_z + \tau_1  \sigma_x\Delta_\perp(x) ~,
\label{eq:H0}
\\
H_1=     \tau_1\frac{ \Delta_\parallel(x)}{p_F} \left(\sigma_y  p_y+\sigma_z  p_z \right)\,,
\label{eq:H1}
\end{eqnarray}
 where $v_F=p_F/m^*$ is the Fermi velocity. 

For ${\bf p}_\parallel=(p_y,p_z)=0$ one has the Hamiltonian \eqref{eq:H0}. This Hamiltonian is supersymmetric, where $\Delta_\perp(x)$ serves as superpotential, since it changes sign across the interface. That is why \eqref{eq:H0} has eigenstates with exactly zero energy. There are two solutions with $E({\bf p}_\parallel=0)=0$, which correspond to different orientations of spin:
 \begin{eqnarray}
\Psi_+(x) \propto  \left(
\begin{matrix}
 1\cr
i\cr
\end{matrix} 
\right)_\tau
 \left(
\begin{matrix}
 1\cr
0\cr
\end{matrix} 
\right)_\sigma
 \exp\left(-\frac{1}{v_F} \int_0^x dx' \Delta_\perp(x')\right) ~,
 \label{eq:ZeroEnergy+}
 \\
 \Psi_-(x) \propto  \left(
\begin{matrix}
 1\cr
-i\cr
\end{matrix} 
\right)_\tau
 \left(
\begin{matrix}
 0\cr
1\cr
\end{matrix} 
\right)_\sigma
 \exp\left(-\frac{1}{v_F} \int_0^x dx' \Delta_\perp(x')\right) \,.
\label{eq:ZeroEnergy_}
\end{eqnarray}

For nonzero ${\bf p}_\parallel$  one may use the perturbation theory with $H_1$ as perturbation, if $|{\bf p}_\parallel|\ll p_F$.
The second order secular equation for the matrix elements of $H_1$ gives the relativistic spectrum $E^2=c^2(p_x^2+p_y^2)$ of the gapless edge states in \eqref{eq:ModesH} with the ``speed of light'':
 \begin{equation}
c=\frac{ \int_0^\infty dx \frac{\Delta_\parallel(x)}{p_F} \exp\left(- \frac{2}{v_F} \int_0^x dx' \Delta_\perp(x')\right)}
{ \int_0^\infty dx   \exp\left(- \frac{2}{v_F} \int_0^x dx' \Delta_\perp(x')\right)} 
 \,.
\label{eq:SpeedOfLight}
\end{equation}
The speed of light of surface fermions is sensitive to the strtucture of the surface layer, and only in the limit of the infinitely thin surface layer, when $\Delta_\parallel(x)=\Delta_\perp(x)=\Delta_B \Theta(x)$,
where $\Theta(x)$ is the Heaviside step function, it approaches the value determined by the bulk order parameter,
$c \rightarrow c_B= \Delta_B/p_F$.

\subsection{The higher order topology}

In applied magnetic field the Pauli term violates the $K$ symmetry of the Hamiltonian, the topological invariant $N_K$, which describes the topology of bulk state, ceazes to exist, the masslesseness of the edge states is not protected any more and they acquire mass\cite{Volovik2010b} (see also Refs.\cite{Mizushima2015,Mizushima2016}):
\begin{equation}
E^2=c^2(p_x^2+p_y^2) +M^2 ~~,~~|M|= \frac{1}{2}\gamma H\,.
\label{eq:EnergyInH}
\end{equation}
However, the topology does not disappear completely, but now it describes the gapped topological surface states.  The line on the surface, which separates the surface domains with different surface topology (different sign of mass $M$), contains $1+1$ gapless fermions, which are protected by combined action of symmetry and topology on the surface\cite{Volovik2010b}. This effect, when the bulk topology disappears, but the surface topology emerges, is called the second order topology, or in general the higher order topology \cite{Trifunovic2019}.

The topologically protected Majorana edge states are under intensive investigations in superfluid $^3$He-B experiments. They are probed through anomalous transverse sound attenuation,\cite{NagaiImpedance,ImpedanceExp1,ImpedanceExp2,ImpedanceExp3}
in measurements of the surface contribution to specific heat,\cite{SpecificHeat,Bunkov2014,Bunkov2015}
and by magnon BEC technique in NMR experiments.\cite{Heikkinen2014}

\subsection{Caroli-de Gennes-Matricon bound states in the vortex core}

The topological properties of the fermionic bound states in the core of the topological objects is determined both by the
real space topology of the object and by the momentum space topology of the environment.

\begin{figure}
 \includegraphics[width=0.5\textwidth]{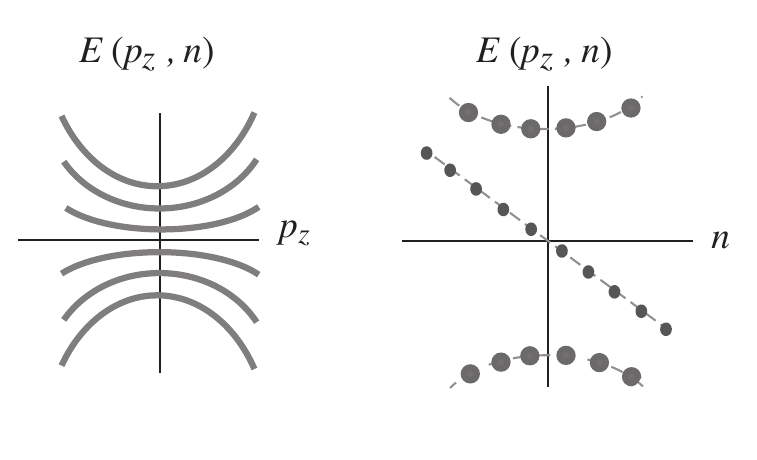}
 \caption{ Illustration of the spectrum of fermion bound states on symmetric singly-quantized (${\cal N}=1)$ vortex in the non-topological $s$-wave superconductors. {\it left}: There are no states with exactly zero energy. {\it right}: However, if one considers the angular momentum quantum number $n$ as continuous variable, there is a branch which crosses zero as a function of $n$. This is the consequence of the special index theorem,\cite{Volovik1993a} which becomes applicable in the semiclassical limit $p_F \gg 1/\xi$  (or which is the same  $\Delta\ll \mu$). 
 }
 \label{CoreStatesS}
\end{figure}

Let us start with the spectrum of the low-energy bound states in the
core
of the vortex with winding number ${\cal N}= 1$ in the isotropic
model of
$s$-wave superconductor. The spectrum is characterized by two quantum numbers, the linear momentum  $p_z$ along the vortex line, and the integer quantum number $n$, which determines the angular momentum $L_z$. 
This spectrum has been analytically obtained in microscopic theory by Caroli,
de
Gennes and Matricon,
\cite{Caroli} see Fig. \ref{CoreStatesS}. The low energy branch has the form:
\begin{equation}
E_n=-\left(n+{1\over 2}\right)\omega_0(p_z) \,.
 \label{Caroli}
\end{equation}
This spectrum is two-fold degenerate due to spin degrees of freedom.
For the nonrelativistic vortex in $s$-wave superconductors the exact zero energy states are absent, see Fig. \ref{CoreStatesS}.
However, typically the level spacing -- the so called minigap -- is very small compared to the energy gap of the
quasiparticles
outside the core, $\omega_0\sim \Delta^2/\mu\ll\Delta$. 
In the approximation, when the minigap is neglected, which is valid in many applications, the angular momentum quantum number $n$ can be considered as
 continuous variable. Then the Eq.(\ref{Caroli}) suggests that there is a branch $E_n$, which as a function of continuous $n$ crosses zero energy level in Fig. \ref{CoreStatesS} ({\it right}).
And, indeed, there is the topological index theorem, which relates the number of branches, which cross zero as a function of $L_z$ and the winding number ${\cal N}$ of the vortex.\cite{Volovik1993a}
This is the analog of index theorem for relativistic cosmic strings, where the theorem relates the number of branches, which cross zero as a function of $p_z$, and the string winding number.\cite{JackiwRossi1981}

\subsection{Flat band of Majorana fermions in the singular vortex in $^3$He-A}
\label{FlatBandVortex}

\begin{figure}
 \includegraphics[width=0.5\textwidth]{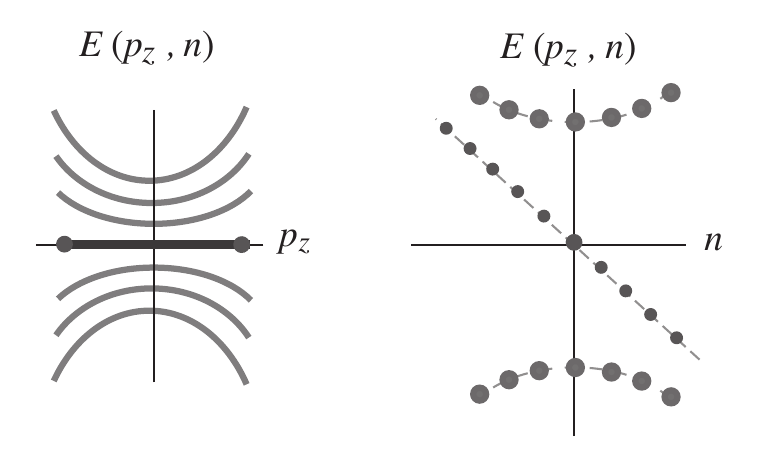}
 \caption{ Illustration of the spectrum of fermion bound states on the most symmetric ${\cal N}=1$ vortex in the chiral Weyl superfluid $^3$He-A. As distinct from Fig. \ref{CoreStatesS}, the spectrum in this topological superfluid contains the states with exactly zero energy at $n=0$ in Eq.(\ref{chiral}). These states form the flat band, which terminates on the projections of two Weyl points to the vortex line, see Fig.\ref{VortexFlatBand}. 
 }
 \label{CoreStatesA}
\end{figure}

Now let us consider fermions in the core of singular ${\cal N}=1$ in  the chiral superfluid $^3$He-A.
The anomalous branch, which crosses zero as function of continuous angular momentum quantum number $n$, remains the same,
see Fig. \ref{CoreStatesA} ({\it right}). The existence of the anomalous branch $E(n)$  depends only on the winding number ${\cal N}$ of the vortex, and is not sensitive to the topology of the bulk superfluid. The latter determines the fine structure of the spectrum. Now the spectrum contains the exact zero energy states:\cite{Volovik1999,Volovik2003}  
\begin{equation}
E_n=-n\omega_0(p_z)\,.
 \label{chiral}
\end{equation}
This is the result of the bulk-defect correspondence: the odd winding number of the phase of the gap function, $\Delta({\bf p}) \propto (p_x + ip_y)$, in bulk is responsible for the zero-energy states in the core.  In  the two-dimensional case the $n=0$ levels represent two spin-denenerate Majorana modes.\cite{ReadGreen2000,Ivanov2001}  The 2D half-quantum vortex, which is the vortex in one spin component, contains single Majorana mode.

\begin{figure}
 \includegraphics[width=0.5\textwidth]{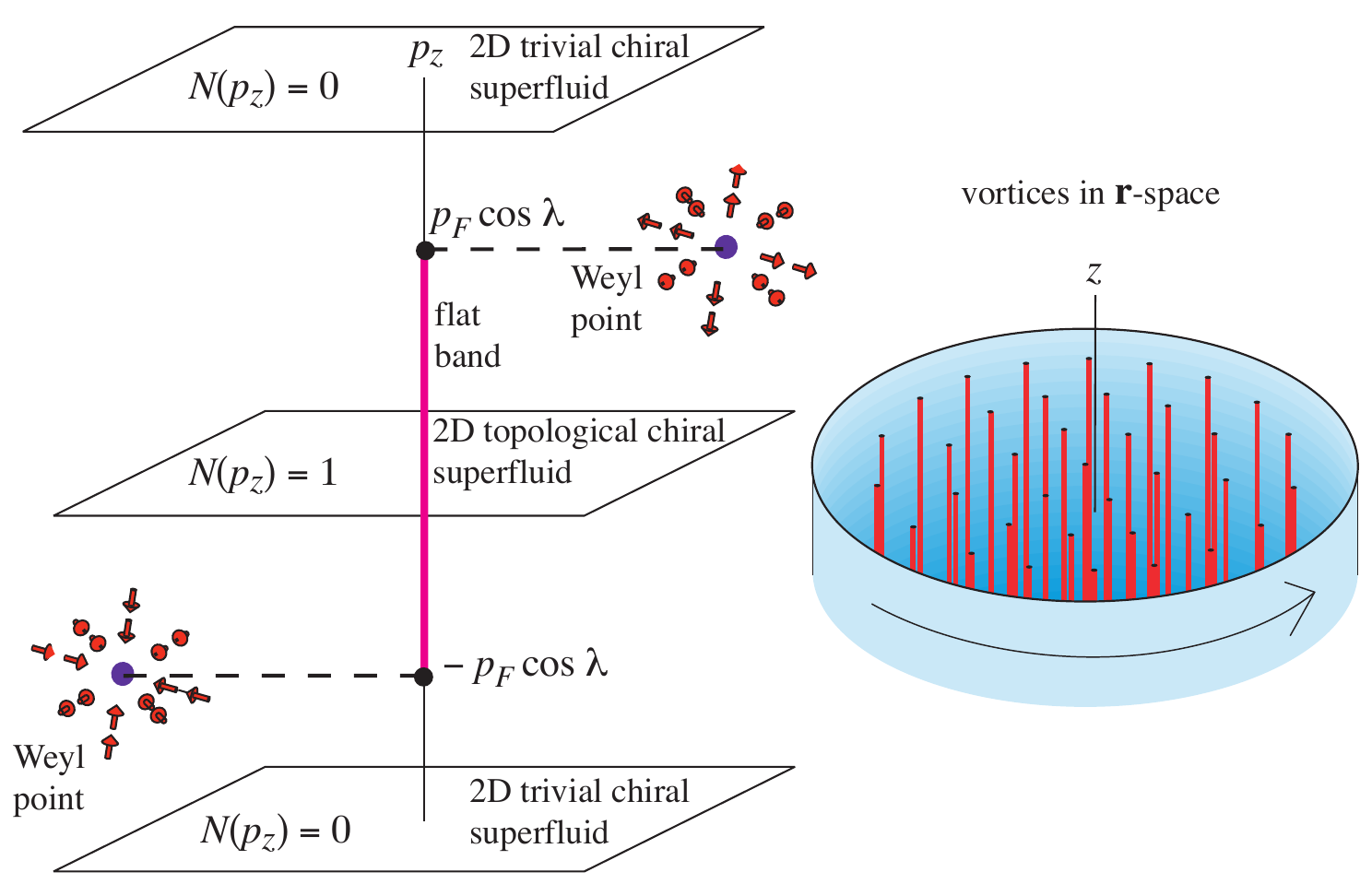}
 \caption{ 
Projections of Weyl points on the direction of the vortex axis (the $z$-axis) determine the boundaries of the flat band
  in the vortex core in Fig. \ref{CoreStatesA}. Weyl point in 3D systems represents the hedgehog (Berry phase monopole) in momentum space. For each plane $p_z={\rm const}$ one has the
effective 2D system with the fully gapped energy spectrum
$E_{p_z}(p_x,p_y)$, except for the planes with $p_{z\pm}=\pm p_F
\cos\lambda$, where the energy $E_{p_z}(p_x,p_y)$ has a node due
to the presence of the hedgehogs in these planes. Topological
invariant $N(p_z)$ in \eqref{TopInvariant2D} is non-zero for $|p_z| < p_F
|\cos\lambda|$, which means that for any value of the parameter
$p_z$  in this interval the system behaves as  2D fully gapped topological superfluid. Point vortex
in such 2D superfluids has fermionic state with exactly zero
energy. For the vortex line in the original 3D system with two Weyl
points this corresponds to the dispersionless spectrum of fermion
zero modes in the whole interval $|p_z| < p_F |\cos\lambda|$
(thick line). This consideration can be extended to the boundary states.
Due to the bulk-boundary correspondence, the 1D boundary of the 2D topological 
insulator with $N(p_z)=1$ contains the branch
 of the spectrum of edge states, 
which crosses zero energy.\cite{Volovik1997} In the 3D system these zeroes form the 1D line of the edge states with zero energy -- the Fermi arc, which is terminated by the projections of the Weyl points to the surface.\cite{Burkov2011a,Burkov2011b}  
 }
 \label{VortexFlatBand}
\end{figure}

In the 3D case the Eq.(\ref{chiral}) at $n=0$ describes the flat band:\cite{KopninSalomaa1991} all the states in the interval $-p_F<p_z<p_F$ have zero energy, where ${\bf K}^{(a]}=\pm p_F\hat{\bf z}$
mark the positions of two Weyl points in the bulk material.\cite{Volovik2011}
The reason for that can be explained using the general case, when the vortex is along the $z$-axis and the Weyl points are at ${\bf K}^{(a]}=\pm p_F\hat{\bf l}$ with $\hat{\bf l}=\hat{\bf z} \cos\lambda + \hat{\bf x}\sin\lambda$,  see Fig. \ref{VortexFlatBand}.
At each $p_z$ except for $|p_z|=p_F\cos\lambda$, i.e. away from the Weyl nodes, the spectrum is gapped, and thus the system represents the set of the fully gapped (2+1)-dimensional chiral superfluids. Such superfluids are characterized of the topological invariant $N(p_z)$, obtained by dimensional reduction from the invariant $N$ in 
Eq.(\ref{MasslessTopInvariant3D}), which desribes the Weyl points:
 \begin{equation}
N(p_z) = \frac{e_{\beta\mu\nu}}{24\pi^2}~
{\bf tr}\int dp_x dp_y d\omega
~ G\partial_{p_\beta} G^{-1}
G\partial_{p_\mu} G^{-1} G\partial_{p_\nu}  G^{-1}\,.
\label{TopInvariant2D}
\end{equation}
Here $p_\mu=(\omega,p_x,p_y)$. Such invariant is responsible for the quantized value of the Hall conductivity  in the absence of external magnetic field  in the (2+1) topological materials with broken time reversal symmetry.\cite{So1985,IshikawaMatsuyama1986,IshikawaMatsuyama1987,Volovik1988a,Haldane1988,Volovik1988b,VolovikYakovenko1989,Yakovenko1989,Golterman1993} 
It is the generalization of the topological invariant introduced for quantum Hall effect.\cite{NiuThoulessWu1985} 

This invariant, which is applicable both to interacting and non-interacting systems, gives 
\begin{eqnarray}
N_3(p_z)=1~~,~~|p_z| < p_F\cos \lambda
 \,,
\label{TopInsulator}
\\
 N_3(p_z)=0~~,~~|p_z| > p_F\cos \lambda
 \,.
\label{Non-TopInsulator}
\end{eqnarray} 
 Since the vortex core of the topologically nontrivial (2+1) superfluid contains the zero energy Majorana mode, one obtains the zero energy states in the whole interval $-p_F\cos\lambda<p_z <p_F\cos\lambda$:
\begin{equation}
E(p_z)=0~~,~~ -p_F\cos\lambda<p_z <p_F\cos\lambda
\,.
\label{FlatBand}
\end{equation}
This is the topological origin of the flat band in the core of the singular $^3$He-A vortex, first calculated in Ref.\cite{KopninSalomaa1991}.

Flat band has been also discussed for vortices in $d$-wave superconductors.\cite{LeeSchnyder2016}

\subsection{Fermion condensation on vortices in polar phase}
\label{FermionCondensation}

\begin{figure}
 \includegraphics[width=0.3\textwidth]{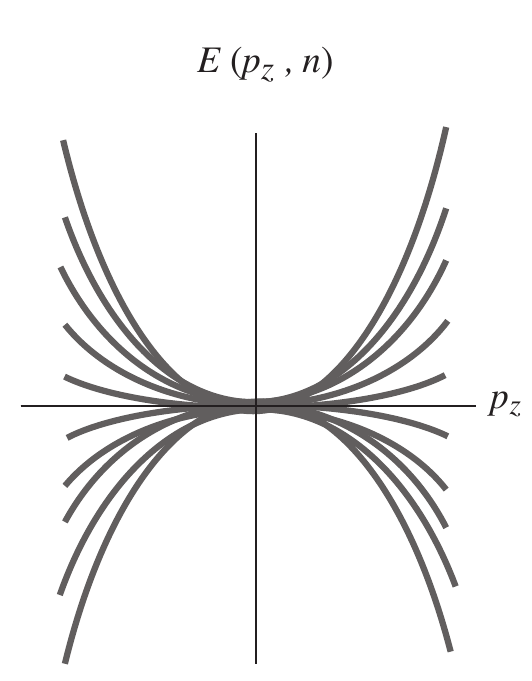}
 \caption{Illustration of the spectrum of fermion zero modes at $|p_z|\ll p_F$ on vortices in the polar phase of superfluid $^3$He. The branches with different $n$ approach zero-energy level at $p_z\rightarrow 0$. This is the consequence of the Dirac nodal ring in the spectrum of the polar phase.
 }
 \label{VortexFermions}
\end{figure}

Here we consider vortices, in which the minigap $\omega_0(p_z)$ vanishes at $p_z=0$. This leads to the enhanced density of states of the fermions in the vortex core, and as a consequence to the non-analytic behavior 
of the DoS as a function of magnetic field in superconductor or of rotation velocity in superfluid.

Examples are provided by vortices\cite{Autti2016}  in the polar phase of superfluid $^3$He.
In general, the minigap for the Caroli-de Gennes-Matricon bound states in the core of symmetric vortices in which the gap function behaves as $\Delta(p_z)f(r)$ is given by the following equation (see e.g. Ref. \cite{Volovik1999}):
\begin{equation}
\omega_0(p_z)  =\frac {
\int_{0}^\infty dr |\psi_0(r)|^2 
\frac{\Delta(p_z)f(r)}{qr }
}
{\int_0^\infty dr
|\psi_0(r)|^2}
~~\,,~~q=\sqrt{p_F^2-p_z^2}\,,
 \label{QuasiclasicalEnergy}
\end{equation}
where
\begin{equation}
\psi_0(r)=
\exp{
\left(-\int^r_0 dr' 
\frac{\Delta(p_z)f(r)}{v_F}\right)}
\,,
 \label{WaveFunction}
\end{equation}
is the wave function of the bound state.

For $s$-wave superconductors one has isotropic gap, $\Delta(p_z)=\Delta_S$; for $^3$He-A the gap has point nodes, $\Delta(p_z)=\Delta_A |{\bf p}_\perp|/p_F= \Delta_A  \sqrt{p_F^2-p_z^2}/p_F$; and for the polar phase in Eq.(\ref{PolarHamiltonian}) the gap has nodal line, $\Delta(p_z)=\Delta_P |p_z|/p_F$.
The nodal line in the polar phase leads to the large suppression of the minigap at small $p_z\ll p_F$:
\begin{equation}
 \omega_0(p_z) =  \omega_{00}
\frac{p_z^2}{p_F^2}\ln\frac{p_F^2}{p_z^2} ~~,~~ \omega_{00}\sim \frac{\Delta_0^2}{E_F}
\,,
 \label{spectrum}
\end{equation}
where $\omega_{00}$ has an order of the minigap in the conventional $s$-wave superconductors.
The spectrum is shown in Fig. \ref{VortexFermions}. All the branches with different $n$ touch the zero energy level. It looks as the flat band in terms of $n$ for $p_z=0$. However, at $p_z\rightarrow 0$ the size of the bound state  wave function diverges, the state merges with the bulk spectrum and disappears.

The effect of squeezing of all energy levels $n$ towards the zero energy
at $p_z\rightarrow 0$ can be called the condensation of Andreev-Majorana
fermions in the vortex core.
The condensation leads to the divergent density of states (DoS) at small energy.  In the vortex cluster with the vortex density density $n_V$ the DoS is
\begin{equation}
N_V=n_V\sum_n \int \frac{dp_z}{2\pi} \delta(\omega - (n+1/2) \omega_0(p_z) ) \,.
 \label{DOS}
\end{equation}
In calculation of Eq.(\ref{DOS}) we assume that the relevant values of $n$ are large, and instead of summation over $n$ one can use the integration over $dn$:
\begin{equation}
N_V = n_V \int \frac{dp_z}{2\pi} \frac{1}{\omega_0(p_z)}\,.
 \label{DOS2}
\end{equation}
According to Eq.(\ref{spectrum}) the integral in Eq.(\ref{DOS2}) diverges at small $p_z$. The infrared cut-off is provided by the intervortex distance $r_V=n_V^{-1/2}$: the size of the wave function of the bound state $\xi p_F/|p_z|$ approaches the intervortex distance when  $|p_z|\sim p_F\xi/r_V$. This cut-off leads to the following dependence of DoS on the intervortex distance:
\begin{equation}
N_V \sim \frac{p_F^2}{\Delta_0r_V}\,.
 \label{DOS3}
\end{equation}
The result in Eq.(\ref{DOS3}) is by the factor $r_V/\xi$ larger than the DoS of fermions bound to conventional vortices.  Since in the vortex array 
$r_V \propto \Omega^{-1/2}$, the DoS has the non-analytic dependence on rotation velocity, $N_V\propto \Omega^{1/2}$. Similar effect leads to the $\sqrt{B}$ dependence of the DoS on magnetic field in cuprate superconductors.\cite{Volovik1993,Moler1997}

\begin{figure}
 \includegraphics[width=0.5\textwidth]{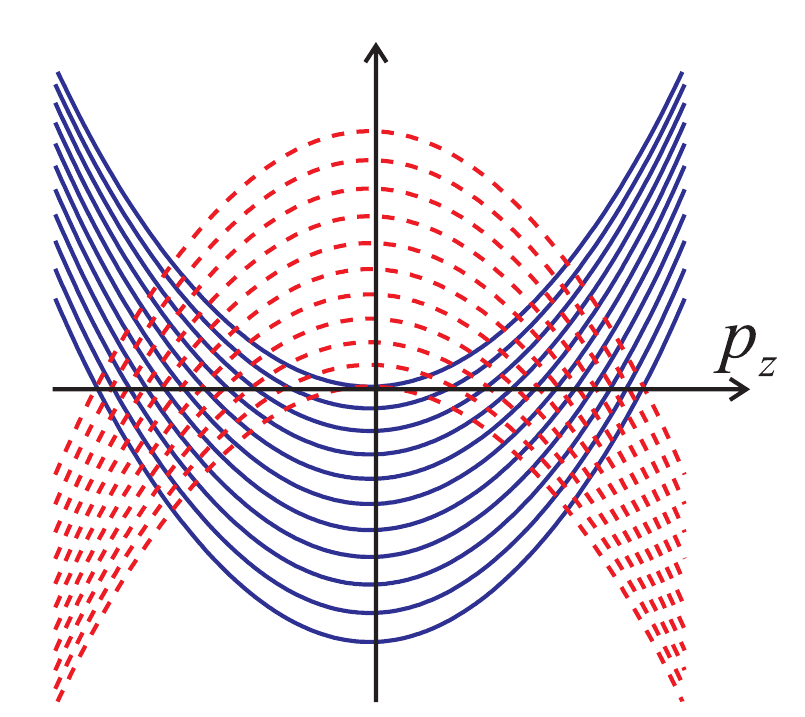}
 \caption{Spectrum of fermions in the axisymmetric $^3$He-B vortex with the A-phase core in Fig. \ref{Evolution}. There are many non-topological branches of spectrum, which
cross zero energy as function of $p_z$ and form
 the one-dimensional Fermi surfaces. The number of such branches $\sim E_F/\Delta_B \gg 1$
 }
 \label{v-vortex}
\end{figure}

\subsection{Vortices with multiple branches crossing zero in $^3$He-B}
\label{multple}

Finally let us mention, that vortices with broken symmetry in the vortex core may contain a large number of the
non-topological branches of spectrum, which cross zero as function of $p_z$,\cite{Volovik1989d,Volovik1991b,Silaev2009,SilaevVolovik2014} see e.g. Fig. \ref{v-vortex} for the spectrum in the $^3$He-B vortex with the A-phase core.

\section{Conclusion}

 At the moment the known phases of liquid $^3$He belong to 4 different topological classes.

(i) The normal liquid $^3$He belongs to the class of systems with topologically protected 
Fermi surfaces. The Fermi surface is described by the first odd Chern number in terms of the Green's function in Eq.(\ref{InvariantForFS}).\cite{Volovik2003}

(ii) Superfluid $^3$He-A and  $^3$He-A$_1$ are chiral superfluids with the Majorana-Weyl fermions in bulk,
which are protected by the Chern number in Eq.(\ref{MasslessTopInvariant3D}). In the relativistic quantum field theories, Weyl fermions give rise to the effect of chiral (axial) anomaly. The direct analog of this effect has been experimentally demonstrated in $^3$He-A.\cite{BevanNature1997} 
It is the first condensed matter, where the chiral anomaly effect has been observed.  

The singly quantized vortices in superfluids with Weyl points contain dispersionless band (flat band) of Andreev-Majorana fermions in their cores.\cite{KopninSalomaa1991}

New phases of the  chiral superfluids have been observed in aerogel. These are the so-called Larkin-Imry-Ma states -- the glass states of the $\hat{\bf l}$-field,\cite{Dmitriev2010} which can be represented as disordered tangle of vortex skyrmions. 
These are the first representatives of the inhomogeneous disordered ground state of the topological material. The spin glass state in the $\hat{\bf d}$-field has been also observed. The recently observed chiral superfluid with polar distortion\cite{Dmitriev2014} also has Weyl points in the spectrum.

(iii) $^3$He-B is the purest example of a fully gapped superfluid with topologically protected
 gapless Majorana fermions on the surface.\cite{Mizushima2015}

(iv) The most recently discovered polar phase of superfluid $^3$He belongs to the class of fermionic materials with topologically protected lines of nodes, and thus contains two-dimensional flat band of Andreev-Majorana fermions on the surface of the sample,\cite{SchnyderRyu2011} see also recent reviews  on superconductors with topologically protected nodes.\cite{SchnyderBrydon2015}

It is possible that with the properly engineered nanostructural confinement one may reach also new topological phases of liquid $^3$He including:

(i) The planar phase, which is the non-chiral superfluid with Dirac nodes in the bulk
and with Fermi arc of Anfreev-Majorana fermions on the surface.\cite{MakhlinSilaevVolovik2014} 

(ii) The two-dimensional topological states in the ultra-thin film, including inhomogeneous phases of superfluid $^3$He films.\cite{Vorontsov2007}   The films with the $^3$He-A and the planar phase order parameters belong to the 2D fully gapped topological materials, which experience the quantum Hall effect and the spin quantum Hall effect in the absence of magnetic field.\cite{VolovikYakovenko1989}

(iii) $\alpha$-state, which contains 4 left and 4 right Weyl points 
in the vertices of a cube, ${\bf K}^a= \frac{p_F}{\sqrt{3}}(\pm \hat{\bf x} \pm \hat{\bf y}\pm \hat{\bf z})$.\cite{VolovikGorkov1985,KlinkhamerVolovik2005} This is close to the high energy physics model with
8 left-handed and 8 right-handed Weyl fermions in the vertices of a four-dimensional cube.\cite{Creutz2008,Creutz2014}
This is one of many examples when the topologically protected nodes in the spectrum serve as an inspiration
for the construction of the relativisitc quantum field theories.

See also the other proposals in Refs.\cite{WimanSauls2015,WimanSauls2018,Levitin2019}.
 
We did not touch the wide area of the bosonic collective modes in the topological superfluids.
Superfluid phases of $^3$He are the objects of the quantum field theory, in which the fermionic quantum fields interact with propagating bosonic modes, some of which have the relativistic spectrum. Among these modes there are analogs of gravitational and electromagnetic fields, $W$ and $Z$ gauge bosons and Higgs fields. Higgs bosons -- the amplitude modes -- have been experimentally investigated
in superfluid $^3$He for many years. For example, among 18 collective modes of $3\times 3$ complex order parameter in $^3$He-B, four are gapless Nambu-Goldstone modes: oscillations of the phase $\Phi$ 
represent the sound waves, while and oscillations of the rotation matrix $R_{\alpha i}$
are spin waves. The other 14 modes are the Higgs modes with energy gaps
of the order of $\Delta_B$. These heavy Higgs modes in $^3$He-B have been investigated both 
theoretically\cite{Vdovin1963, Maki1974, Nagai1975,Tewordt-Einzel1976} and 
experimentally.\cite{Giannetta1980, Mast1980,Avenel1980, Lee1988, Collett2012} 
Due to spin-orbit interaction one of the spin-wave Goldstone mode acquires a small mass and becomes the light Higgs boson.
The properties of the heavy and light Higgs modes in $^3$He-A, $^3$He-B and in the polar phase suggest different scenarios for the formation of the composite
Higgs bosons in particle 
physics.\cite{VolovikZubkov2013,VolovikZubkov2014,VolovikZubkov2015,Zavjalov2016,Sauls2015,Lee2016}

One may expect the topological confinement  of topological objects of different dimensions in real space, momentum space and in the combined phase space. Examples are: Fermi surface with Berry phase flux in $^3$He-A in the presence of the superflow;\cite{Volovik2003}
 the confinement of the point defect (monopole) and the line defect (string) in real space;\cite{Cornwall1999} the nexus in momentum space,\cite{Mikitik2006,TeroNexus2015}  which combines the Dirac lines, Fermi surfaces and the crossing points; the so-called type II Weyl point, which leads to formation of the analog of the black hole horizon \cite{VolovikZubkov2014b,Soluyanov2016,YongXu2015,VolovikZhang2017,VolovikUFN2018}; etc.

Topology also gives rise to different types of superfluid glass states, including skyrmion glass, Weyl glass, analogs of spin foam, etc. \cite{Glasses}

{\bf Acknowledgements}. This work has been supported by the European Research Council (ERC) under the European Union's Horizon 2020 research and innovation programme (Grant Agreement No. 694248).

\end{document}